\documentclass[english, 10 pt]{amsart}                                 
\usepackage{babel}                                                          
\usepackage{amsthm}
\usepackage{amsmath}                                                         
\usepackage{epsfig}                                                          
\numberwithin{equation}{section}                                             
\numberwithin{figure}{section}                                               
\begin{document}                                                             
\baselineskip 0.2truecm                                                      
\renewcommand{\dfrac}{\displaystyle\frac}                                    
\newcommand{\disp}{\displaystyle}                                    
\newcommand{\mbf}{\mathbf}
\newcommand{\dint}{\displaystyle\int}
\renewcommand{\u}{\underline}                                        
\renewcommand{\o}{\overline}                                        
\newtheorem{thm}{Theorem}     [section]                                    
\newtheorem{definition}{Def.}   [section]                                    
\newtheorem{obs}{Remark}  [section]
\title[Suspension flows  in  pipelines with partial phase  separation]
{Suspension flows    in  pipeline  with     partial phase   separation (*)
\footnote{\tiny (*) Work partially  supported  by the Italian  CNR
Strategic  project on Fluid Dynamics and  by the Italian MURST Project
on Mathematical Analysis of Phase Transitions}}
\author[Alessandro Speranza]{Alessandro Speranza\\\vskip 0.3cm
  {\emph Dip.Matematica ``U.Dini'',   
    Universit\`a degli Studi di Firenze\\                                  
    v.le Morgagni 67/a, 50134 Firenze, Italy}\\
  {\tt alessandro.speranza@math.unifi.it}}
\baselineskip .70truecm                                                     
\maketitle
\begin{abstract}
The  formulation of a  model for the evolution of the
flow of a  solid-liquid mixture (coal-water) in  a horizontal pipeline
with partial  phase separation is the  aim of this  work.  Problems of
instabilities due to complex eigenvalues, observed in previous models,
seem to  be completely solved in  the present model, in  which we give
the genesis  of the different  terms written in the  equations, coming
from  the natural  definition of  mass and  momentum balance,  and the
consequent  proof of well-posedness  of the  obtained PDE  system with
boundary-Cauchy data.

The model   describes a three-layer   flow. Most of  the  material is
carried by  the  upper layer, while  the  bottom layer consists of  an
immobile sediment. The intermediate layer grows to a maximum thickness
and has   the role of regulating  the mass  exchange between   the extreme
layers.

In  the last  section we  present  some simulations  for a  particular
choice of  flow regime, and boundary-Cauchy data,  that were suggested
by experimental results provided by Snamprogetti (Fano, Italy).
\end{abstract}
\vskip 0.4cm
\centerline{\tiny{{\bf keywords:} suspensions, phase-separation,
turbulent flow, hyperbolic, well-posedness}}                                 

\section{The physical problem}\label{sec:introduction}

In this paper  we formulate  a mathematical model   for the flow in  a
horizontal pipeline of a solid-water mixture  with a progressive phase
separation due to gravity.   

The specific case we have in  mind is the one of ``dilute'' coal-water
suspensions and was suggested to us by Snamprogetti (Fano, Italy). The
fresh mixture has a coal concentration of about $50\%$ (in weight) and
a particle size distribution centered  at $0.205 mm$ ($\leq 1.25 mm$).
The rheological properties of such a system are totally different from
the ones of coal-water slurries  that have been studied extensively in
a  number of papers  (\cite{comparini-de-angelis}, \cite{fasano-ecmi},
\cite{fasano-primicerio-2},                   \cite{fasano-primicerio},
\cite{primicerio}, \cite{shook}, \cite{persol}).

Indeed slurries, which  can have coal concentrations up  to $70\%$ and
are prepared using smaller particles, are stable at rest thanks to the
action of  chemical additives and exhibit  partial sedimentation under
stress,  mainly because  of the  presence of  impurities that  are not
affected by  the additive  and are no  longer sustained by  the slurry
yield stress in dynamical conditions.

The situation  here is completely  different,  because the  liquid and
solid components tend to  separate under gravity both  at rest, and in
dynamical conditions.

The approach we  are going to follow is  in the typical framework  of
stratified flows: in any transversal cross section of the pipe we have
three layers, whose motion  and composition are described by  averaged
quantities.

The upper layer  carries most of the material,  while the bottom layer
is immobile (a stationary deposit that should be avoided in practice).
The middle layer has a  density intermediate between the lighter, fast
flowing,  upper phase,  and  the  sediment.  We  suppose  that at  the
beginning just  one phase  exists, having the  same properties  of the
material entering  the pipeline. Therefore  the system goes  through a
first  stage in which  there is  no immobile  layer, while  the denser
flowing layer grows up to  a thickness $\Delta$, depending on the pipe
discharge. Once this value of the thickness is reached, the stationary
sediment appears and the intermediate layer moves upwards, keeping its
thickness  $\Delta$  and  its  concentration,  and  transmitting  solid
material from the main phase to the sediment.

The idea of  the intermediate layer is   inspired to a  model that has
succesfully explained  sedimentation  in slurries  (see \cite{fasano},
\cite{mancini}), although as we said ,the general properties are quite
different in the two cases.

The  scheme  utilizing  piecewise  constant velocities  over  a  cross
section,  although very  convenient  and frequently  used in  pipeline
modeling, is already largely approximated, so that, taking the density
constant in  each layer, is not  only a reasonable  assumption, but it
looks the only  one really consistent with the  general setting of the
problem.  In fact,  the way density varies, over  a cross section, can
be measured  by means of  a Gamma-Densimeter, however the  accuracy is
not such that one could look for a more sophisticated model.

%"THE WAY DENSITY VARIES OVER A
%CROSS SECTION CAN ME MEASURED BY MEANS OF A GAMMA-DENSIMETER. HOWEVER THE
%ACCURACY IS NOT SUCH THAT ONE COULD LOOK FOR A MORE SOPHISTICATED MODEL.
%INDEED, THE SCHEME UTILIZING PIECEWISE CONSTANT VELOCITIES OVER A CROSS
%SECTION, ALTHOUGH VERY CONVENIENT AND FREQUENTLY USED IN PIPELINE 
%MODELING, IS ALREADY LARGELY APPROXIMATED, SO THAT TAKING THE DENSITY
%CONSTANT IN EACH LAYER IS NOT ONLY A REASONABLE ASSUMPTION, BUT IT LOOKS
%THE ONLY ONE REALLY CONSISTENT WITH THE GENERAL SETTING OF THE PROBLEM." 

Previous models  in the literature  describing flows of  mixtures (see
\cite{multiphase-flow},    \cite{cheng-wang-ling},    \cite{liepmann},
\cite{ramshaw-trapp},    \cite{olga},   \cite{shook},   \cite{shook1},
\cite{shook-book}, \cite{stewart}  ) could not be adapted  to our case
for   different    reasons\footnote{Problems   of   ill-posedness   in
  \cite{olga}  were pointed out  (see Remark  ~\ref{obs:ill-posed}) in
  \cite{cheng-wang-ling}, while  the stationary models  present in the
  literature are not interesting in our case, since stationary flow is
  reached at such large dinstances  from the entrance of the pipeline,
  that  it is  not  observable  in practice.   In  addition, it  would
  correspond to  the situation in  which only water is  transported by
  the main  flow, which is of course  of no practical use.}  , so that
the  model we  are  going to  illustrate is,  as  far as  we know,  an
original  contribution  towards the  study  of non-steady  suspensions
flows with phase separation.
%%% Local Variables: 
%%% mode: latex
%%% TeX-master: t
%%% End: 
%\input{twolayer}
\section{The two-layer flow model} \label{sec:2fasi}

Let us deal first with the initial stage of  the flow in which we have
just two layers.

\begin{figure}[htb]                                                    
  \begin{picture}(0,0)%
\includegraphics{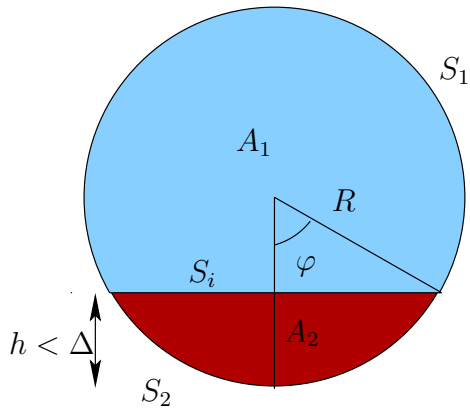}%
\end{picture}%
\setlength{\unitlength}{3947sp}%
\begingroup\makeatletter\ifx\SetFigFont\undefined%
\gdef\SetFigFont#1#2#3#4#5{%
  \reset@font\fontsize{#1}{#2pt}%
  \fontfamily{#3}\fontseries{#4}\fontshape{#5}%
  \selectfont}%
\fi\endgroup%
\begin{picture}(2854,2480)(751,-1636)
\put(751,-1336){\makebox(0,0)[lb]{\smash{\SetFigFont{12}{14.4}{\familydefault}{\mddefault}{\updefault}
$h<\Delta$
}}}
\put(2176,-61){\makebox(0,0)[lb]{\smash{\SetFigFont{12}{14.4}{\familydefault}{\mddefault}{\updefault}$A_1$}}}
\put(3451,389){\makebox(0,0)[lb]{\smash{\SetFigFont{12}{14.4}{\familydefault}{\mddefault}{\updefault}$S_1$}}}
\put(1876,-886){\makebox(0,0)[lb]{\smash{\SetFigFont{12}{14.4}{\familydefault}{\mddefault}{\updefault}$S_i$}}}
\put(2551,-811){\makebox(0,0)[lb]{\smash{\SetFigFont{12}{14.4}{\familydefault}{\mddefault}{\updefault}$\varphi$}}}
\put(1576,-1636){\makebox(0,0)[lb]{\smash{\SetFigFont{12}{14.4}{\familydefault}{\mddefault}{\updefault}$S_2$}}}
\put(2476,-1261){\makebox(0,0)[lb]{\smash{\SetFigFont{12}{14.4}{\familydefault}{\mddefault}{\updefault}$A_2$}}}
\put(2776,-436){\makebox(0,0)[lb]{\smash{\SetFigFont{12}{14.4}{\familydefault}{\mddefault}{\updefault}
$R$
}}}
\end{picture}
\caption{Two-layer flow model, vertical section}                   
  \label{fig:2fasi} 
\end{figure}

We  denote by  $x$ the  coordinate   along the flow.    At any section
$x=const.$,   the two layers  occupy two  regions  of areas $A_1$ (the
upper layer),   $A_2$ (the lower),   separated by  a  straight line of
length $S_i$. The perimeters of the outer boundaries are $S_1$, $S_2$,
respectively. In  the  stage we are considering   the thickness of the
lower layer is less than $\Delta$ (see fig.~\ref{fig:2fasi}).

The  flow   model is   based   on mass  and   momentum conservation
accounting for the exchange of liquid and solid between the two layers.

The basic quantities entering the model are, besides $A_1$, $A_2$,\\

\begin{tabular}{ll}
  $\alpha$: & solid volume fraction in layer 1;\\
  $\beta$: & solid volume fraction in layer 2;\\
  $U$: & velocity of layer 1;\\  
  $V$: & velocity of layer 2.\\
\end{tabular}  
\\
\\
The solid volume transfer rate from layer 1 to layer 2 will be assumed
to  be  $\alpha A_1\psi$,  where $\psi$  is  a positive constant.  The
corresponding  quantity for   the liquid   component can be    written
$(1-\alpha)A_1\varphi$ where $\varphi$  turns out to  be a function of
$\alpha,\beta,\psi$, as we shall see.

\subsection{Mass balance}

Both the components are incompressible, so we have the following volume
conservation laws:
\begin{itemize}                                                       
\item LAYER 1-solid 
  \begin{equation} \label{eq:s1}                                
    \partial_t(\alpha A_1)+\partial_x(\alpha A_1U)=-\alpha A_1\psi    
  \end{equation}                                                      
\item LAYER 1-liquid                                                  
  \begin{equation} \label{eq:l1}                                      
    \partial_t\left[(1-\alpha)A_1\right]+\partial_x\left[(1-\alpha) 
      A_1U\right]=-(1-\alpha) A_1\varphi  
  \end{equation}   
\item LAYER 2-solid
  \begin{equation} \label{eq:s2}
    \partial_t(\beta A_2)+\partial_x(\beta A_2V)=\alpha A_1\psi
  \end{equation}                                                     
\item LAYER 2-liquid
  \begin{equation} \label{eq:l2} 
    \partial_t\left[(1-\beta) A_2\right]+\partial_x\left[(1-\beta)
      A_2V\right]=(1-\alpha) A_1\varphi                   
  \end{equation} 
\end{itemize}                                                         
Clearly $A_1$ and $A_2$ are related by:
\begin{equation}\label{eq:Area}
  A_1+A_2=A
\end{equation}
Also the two velocities $U$,$V$  are not independent, since summing up
equations      \eqref{eq:s1},     \eqref{eq:l1},     \eqref{eq:s2},
\eqref{eq:l2}, we obtain  that the discharge $Q$ along  the pipe does
not depend on  $x$, as a natural consequence  of the incompressibility
of the mixture:
\begin{equation} \label{eq:cons.tot}
  \partial_x(A_1U+A_2V)=0\;,
\end{equation}
so that $Q$ may depend on $t$ only, thus:
\begin{equation} \label{eq:V}
  V=\dfrac{Q(t)-A_1U}{A_2}\;,\; \text{if}\;A_2>0
\end{equation} 
The  above equations are not enough  to describe the evolution of the
bottom  layer.  We   assume that the   solid  concentration  in  it is
constant, i.e.:
\begin{equation}\label{eq:beta_const}
  \beta=const.
\end{equation}
Thus we can divide eq.~\eqref{eq:s2} by  $\beta$ and ~\eqref{eq:l2} by
$(1-\beta)$ and realise that
\begin{equation} \label{eq:varphi(psi)}
  \varphi=\dfrac{\alpha}{1-\alpha}\dfrac{1-\beta}{\beta}\psi 
\end{equation}                                              
and that ~\eqref{eq:s2}, ~\eqref{eq:l2}  represent the  same equation,
as  well as  ~\eqref{eq:l1}  is   a  consequence of    ~\eqref{eq:s1},
~\eqref{eq:s2} and ~\eqref{eq:varphi(psi)}

Therefore we reformulate all the  prevoius conservation laws using two
equations   only in the    three  unknowns $\alpha$, $A_1$,  $U$.  For
instance  we can  replace  $A_2$ and $V$  in ~\eqref{eq:s2}  using the
expressions ~\eqref{eq:Area}, ~\eqref{eq:V}, obtaining:
\begin{equation} \label{eq:2-bifase}                                 
  \partial_tA_1+\partial_x(A_1U)=-\dfrac{\alpha}{\beta}A_1\psi\;,     
\end{equation}                                                         
and then we can modify ~\eqref{eq:s1} as follows:
\begin{equation} \label{eq:1-bifase}                              
  \partial_t\alpha+U\partial_x\alpha=-\alpha\left(1-
    \dfrac{\alpha}{\beta}\right)\psi
\end{equation}                                                         
We need one more equation which has to be obtained from the balance of
momentum.

Before deriving the  missing equation, we  define the average densities
$\rho_1$, $\rho_2$ in the two layers:
\begin{equation}\label{eq:rho1}
  \rho_1=\alpha\rho_s+(1-\alpha)\rho_w
\end{equation}
\begin{equation}\label{eq:rho2}
  \rho_2=\beta\rho_s+(1-\beta)\rho_w
\end{equation}
($\rho_s$  and $\rho_w$ are  the  densities of  the solid  and of  the
liquid components).

While $\rho_2$  is constant, $\rho_1$  varies. From ~\eqref{eq:s1} and
~\eqref{eq:l1}, we obtain the global volume balance in $A_1$:
\begin{equation} \label{eq:cons1}
  \partial_tA_1+\partial_x\left(A_1U\right)=
  -\dfrac{\alpha}{\beta}A_1\psi
\end{equation}
after using ~\eqref{eq:varphi(psi)}. Similarly we have
\begin{equation} \label{eq:cons2}
  \partial_tA_2+\partial_x\left(A_2V\right)=
  \dfrac{\alpha}{\beta}A_1\psi\;,
\end{equation}
which is nothing but ~\eqref{eq:s2} divided by $\beta$.

Now, multiplying   ~\eqref{eq:s1}  by   $\rho_s$, ~\eqref{eq:l1}    by
$\rho_w$, and adding the results, we get:
\begin{equation}\label{eq:massa1}
  \partial_t\left(\rho_1A_1\right)+\partial_x\left(\rho_1A_1U\right)
  =-A_1\rho_2\dfrac{\alpha}{\beta}\psi\;,
\end{equation}
and since the left hand side is 
\[-\rho_1A_1\dfrac{\alpha}{\beta}\psi+A_1\left
  (\dfrac{\partial\rho_1}{\partial t}+U\dfrac{\partial\rho_1}
  {\partial x}\right)\;,\]
we deduce how $\rho_1$ varies along the flow:
\begin{equation}\label{eq:D1rho1}
  D_1\rho_1=-\dfrac{\alpha}{\beta}\psi\left(\rho_2-\rho_1\right)
\end{equation}
where,  here  and  in  the  following, $D_1$  and  $D_2$,  denote  the
Lagrangian derivatives along the respective flows:
\begin{equation}\label{eq:derivate}
  \begin{array}{lcl}  
    D_1=\dfrac{\partial}{\partial t}+U\dfrac{\partial}{\partial x} 
    & , & D_2=\dfrac{\partial}{\partial t}+V\dfrac{\partial}
    {\partial x}
  \end{array}
\end{equation}
As we are in the case $\alpha<\beta$ and $\rho_s>\rho_w$, and
\begin{equation}\label{eq:rho2-rho1}
  \rho_2-\rho_1=\left(\beta-\alpha\right)\left(\rho_s-\rho_w\right)\,,
\end{equation}
we  find   out that (from   ~\eqref{eq:D1rho1})   $\rho_1$ is strictly
decreasing along the flow, as long as $\alpha>0$.

\subsection{Momentum balance}

Now we write the following pair of momentum balance equations per unit
length:
\begin{itemize}
\item LAYER 1 
  \begin{equation} \label{eq:mom1}
    D_1(A_1\rho_1U)=A_1G-\tau_1S_1\mp\tau_iS_i+UD_1(\rho_1A_1)
  \end{equation}
\item LAYER 2 
  \begin{equation} \label{eq:mom2}
    D_2(A_2\rho_2V)=A_2G-\tau_2S_2\pm\tau_iS_i-\rho_2UD_2A_1
  \end{equation}
\end{itemize}
where  $G$  denotes  the  absolute  value of  the  pressure  gradient,
$\tau_1$, $\tau_2$  are the stresses at the  respective wall portions,
$\tau_i$ is the interfacial stress, with ``+'', in ~\eqref{eq:mom2} if
$U>V$.  We will come back to the form of such stresses later on .

Note  that since  $D_jA_1<0$   for  $j=1,2$ and  $D_1\rho_1<0$  (see
~\eqref{eq:D1rho1}), we have a momentum loss in ~\eqref{eq:mom1} and a
momentum gain in ~\eqref{eq:mom2}.

The genesis of the loss and gain  terms is the following. Consider the
motion  of volume  element $A_1dx$  of the  upper  layer. In  the time
element       $dt$  it     will     transfer    the     mass   element
$|D_1(\rho_1A_1)|dx\,dt$  to the other  layer, with the corresponding
(negative) momentum transfer $UD_1(\rho_1A_1)$,   per unit length  and
unit time.

If we follow the motion of an element $A_2\,dx$ of the lower layer, it
will acquire the  volume $D_2A_2\,dx\,dt$ in  the interval time  $dt$. 
The corresponding mass increase is  obtained multiplying by  $\rho_2$
(constant) the volume  accretion, and by   $U$, in order to  get the
momentum variation. Remember  that the  matter  coming from  the upper
layer, had a flow velocity $U$ and that it is slowed down to $V$ after
passing the interface.

We can rewrite ~\eqref{eq:mom1} and ~\eqref{eq:mom2} in a more expressive way:
\begin{equation}\label{eq:f=ma1}
  \rho_1A_1D_1U=A_1G-\tau_1S_1\mp\tau_iS_i
\end{equation}
\begin{equation}\label{eq:f=ma2}
  \rho_2A_2D_2V=A_2G-\tau_2S_2\pm\tau_iS_i+\rho_2(U-V)D_2A_2
\end{equation}
\begin{obs}
  The  first equation says   that the ``donor''  system obeys Newton's
  law\footnote{Decompose  the  mass    element   $\rho_1A_1\,dx$  into
    $(\rho_1+D_1\rho_1\,dt)(A_1+D_1A_1\,dt)dx$,   which    is still in
    layer    1    after   $dt$    and     the   complementary     part
    $-D_1\rho_1\,dt(A_1+D_1A_1\,dt)dx  -  \rho_1D_1A_1\,dt\,dx$ (which
    is $O(dt)$). The main portion moves according to
    \[(\rho_1+D_1\rho_1\,dt)(A_1+D_1A_1\,dt)D_1U= \text{external 
      forces + interaction with the}\;O(dt)\;\text{portion}\] 
    which gives ~\eqref{eq:f=ma1} as $dt\rightarrow 0$.}.\\ 
  The  second equation  says  that the accepting system  experiences an
  extra  inertia force due to  the relative momentum of the transferred
  element.\\
  Note that,  due  to  the relative  motion  of the  layers, the  mass
  element  $|D_1(\rho_1A_1)|dx\,dt$     is distributed among different
  elements    of  the   lower   layer. Similarly,   the  mass increase
  $\rho_2D_2A_2\,dx\,dt$ is the  sum of the contribution  of different
  elements    of   the  upper   layer.  Therefore   ~\eqref{eq:f=ma1},
  ~\eqref{eq:f=ma2}  should  not be used  to  get the overall momentum
  balance.  
\end{obs}
%%% Local Variables: 
%%% mode: latex
%%% TeX-master: t
%%% End: 
%\input{stress}
\subsubsection{The stress terms}\label{sec:stress} 

In ~\eqref{eq:mom1}, ~\eqref{eq:mom2} we  take account of the wall and
interface shear stress. Generally speaking, we introduce wall-specific
shear stress (see ~\cite{bifase2}):
\begin{equation}\label{eq:tau}
  \tau=\dfrac{\lambda}{2}\rho v^2
\end{equation}
where
\begin{equation}\label{eq:lambda}
  \lambda=c\left(\dfrac{Dv}{\nu}\right)^{-n}
\end{equation}
is the friction factor (see  \cite{landau}, \cite{bifase2}), $v$ is the
phase velocity,  $\rho$  is the  average phase  density,  $D$ is   the
equivalent hydraulic diameter, namely (\cite{bifase2}):\\
\\
\centerline{
  \begin{tabular}{||l|c|c|c||}\hline\hline
    & $v_a>v_b$ & $v_a<v_b$ & $v_a\simeq v_b$\\\hline\hline
    $D_a$ & $\dfrac{4A_a}{S_a+S_i}$ & $\dfrac{4A_a}{S_a}$  & 
    $\dfrac{4A_a}{S_a}$\\\hline
    $D_b$ & $\dfrac{4A_b}{S_b}$ & $\dfrac{4A_b}{S_b+S_i}$  & 
    $\dfrac{4A_b}{S_b}$\\\hline\hline
  \end{tabular}\\}
\\
\\
with usual notations, and the coefficients  $c$, $n$ can be evaluated,
in the case of Newtonian fluids, for stratified flow, depending on the
flow regime (laminar, turbulent) as:\\
\\
\centerline{
  \begin{tabular}{||l|c|c||}\hline\hline
    reg. & $c$ & $n$ \\ \hline\hline
    Lam. & $16$ & $1$ \\ \hline
    Turb.& $0.046$ & $0.2$ \\ \hline\hline
  \end{tabular}\\}
\\
\\
In ~\eqref{eq:lambda}, $\nu$ is the kinematic viscosity:
\begin{equation}\label{eq:visco}
  \nu=\dfrac{\eta}{\rho}
\end{equation} 
In the  expression of the interface  stress, $v$ has to be interpreted
as the relative velocity $(U-V)=W$ of the two layers so that:
\begin{equation}\label{eq:tau_i}
 \tau_i=\dfrac{\lambda_i}{2}\rho_i W^2
\end{equation}
Here $\lambda_i$ and $\rho_i$ are  the friction factor and the density
of the faster phase.

\subsubsection{A model for the viscosity}
\begin{figure}[htb]                                             
  \begin{picture}(0,0)%
\includegraphics{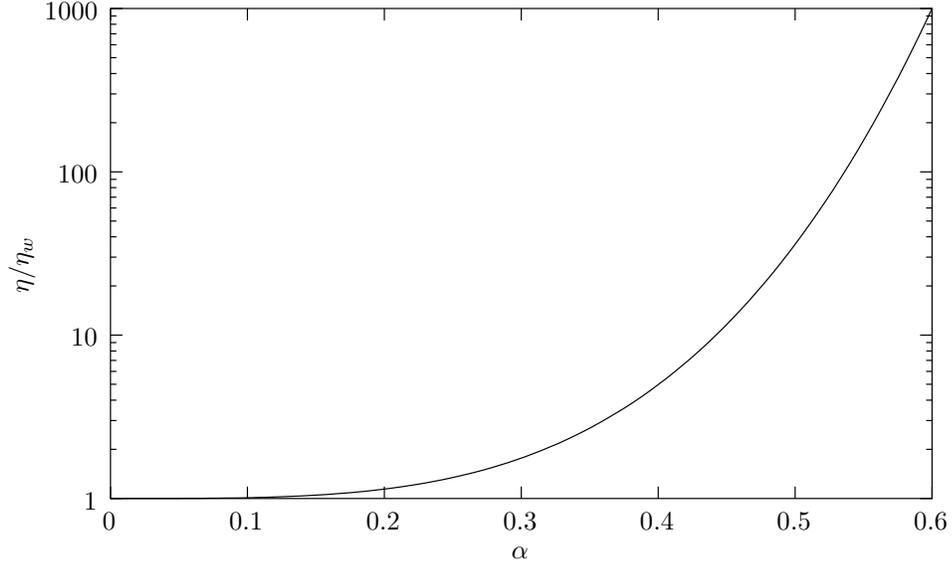}%
\end{picture}%
\setlength{\unitlength}{3947sp}%
\begingroup\makeatletter\ifx\SetFigFont\undefined%
\gdef\SetFigFont#1#2#3#4#5{%
  \reset@font\fontsize{#1}{#2pt}%
  \fontfamily{#3}\fontseries{#4}\fontshape{#5}%
  \selectfont}%
\fi\endgroup%
\begin{picture}(5843,3540)(260,-3106)
\put(857,-2751){\makebox(0,0)[rb]{\smash{\SetFigFont{10}{12.0}{\familydefault}{\mddefault}{\updefault}
1
}}}
\put(857,-1724){\makebox(0,0)[rb]{\smash{\SetFigFont{10}{12.0}{\familydefault}{\mddefault}{\updefault}
10
}}}
\put(857,-698){\makebox(0,0)[rb]{\smash{\SetFigFont{10}{12.0}{\familydefault}{\mddefault}{\updefault}
100
}}}
\put(857,329){\makebox(0,0)[rb]{\smash{\SetFigFont{10}{12.0}{\familydefault}{\mddefault}{\updefault}
1000
}}}
\put(931,-2875){\makebox(0,0)[b]{\smash{\SetFigFont{10}{12.0}{\familydefault}{\mddefault}{\updefault}
0
}}}
\put(1791,-2875){\makebox(0,0)[b]{\smash{\SetFigFont{10}{12.0}{\familydefault}{\mddefault}{\updefault}
0.1
}}}
\put(2651,-2875){\makebox(0,0)[b]{\smash{\SetFigFont{10}{12.0}{\familydefault}{\mddefault}{\updefault}
0.2
}}}
\put(3511,-2875){\makebox(0,0)[b]{\smash{\SetFigFont{10}{12.0}{\familydefault}{\mddefault}{\updefault}
0.3
}}}
\put(4371,-2875){\makebox(0,0)[b]{\smash{\SetFigFont{10}{12.0}{\familydefault}{\mddefault}{\updefault}
0.4
}}}
\put(5231,-2875){\makebox(0,0)[b]{\smash{\SetFigFont{10}{12.0}{\familydefault}{\mddefault}{\updefault}
0.5
}}}
\put(6091,-2875){\makebox(0,0)[b]{\smash{\SetFigFont{10}{12.0}{\familydefault}{\mddefault}{\updefault}
0.6
}}}
\put(425,-1211){\rotatebox{90.0}{\makebox(0,0)[b]{\smash{\SetFigFont{10}{12.0}{\familydefault}{\mddefault}{\updefault}
$\eta/\eta_w$
}}}}
\put(3511,-3061){\makebox(0,0)[b]{\smash{\SetFigFont{10}{12.0}{\familydefault}{\mddefault}{\updefault}
$\alpha$
}}}
\end{picture}
\caption{Exponential model for the viscosity of the mixture}  
  \label{fig:viscosity}                                         
\end{figure}

The viscosity of  the mixture is related to  solid fraction.  For this
reason   we   select  a   particular   behaviour  of   $\eta(\alpha)$,
considering, on one hand, the  typical value of the viscosity of water
($0.01 g  cm^{-1}s^{-1}$), corresponding  to $\alpha=0$, and,  on the
other hand, the value of  the viscosity measured for a suspension with
solid fraction $0.5$, typically $0.35 g cm^{-1}s^{-1}$ (Snamprogetti).
Since  the bottom  layer with  solid fraction  $0.7$, is  at  rest, we
impose that  viscosity increases  considerably beyond the  value $0.5$
for  $\alpha$. These  considerations suggest  the following  choice of
$\eta(\alpha)$:
\begin{equation}
  \eta(\alpha)=\eta_w\left(\dfrac{\eta_2}{\eta_w}\right)
  ^{\left(\dfrac{\alpha}{\beta}\right)^b}
\end{equation}
(see  Fig. ~\ref{fig:viscosity})  where $\eta_w$  is the  viscosity of
water  and $b$ is  a parameter  to be  determined in  such a  way that
$\dfrac{\eta}{\eta_w}=35$,   for   $\alpha=0.5$,   and   $10^3$,   for
$\alpha=\beta\,(=0.6)$.
%For our aims, which is to  develope a simulation for separation in two
%layers in motion with solid fraction $\alpha\leq 0.5$ and $\beta=0.6$,
%it  is  enough if  we  give the  behaviour  of  $\eta(\alpha)$ in  the
%interval $[0:0.5]$, for the upper  layer, and a constant value for the
%middle layer, as we said, taking into account the fact that, for solid
%fractions  more than 0.6,  like the  bottom layer  ($\gamma=0.7$), the
%matter  is   stationary.   For  this  reason  we   put,  for  instance
%$\eta(\beta)=10 gcm^{-1}s^{-1}$ and
%\begin{equation}\label{eq:viscosity}
%  \eta(\alpha)=\eta_we^{c\alpha}\;\;\text{for}\;\;\alpha\leq 0.5
%\end{equation}
%where
%\begin{equation}\label{eq:param}
%  c=2\log\left(\dfrac{0.35}{0.01}\right)
%\end{equation}
\begin{obs}
  The selection above of  $\eta(\alpha)$ is extrapolated from just two
  experimental  data,  so  that  it  can  be used  to  obtain  just  a
  qualitative, although quite realistic, description of the system.
\end{obs}
%Putting  all these considerations
%together, we end up with a function
%\begin{equation}
%  \left(\dfrac{\eta_2}{\eta_w}\right)
%  ^{\left(\dfrac{\alpha}{\beta}\right)^b}
%\end{equation}
%where $\eta_w$ is the viscosity of  water and $b$ is a paramater to be
%determined, imposing the  value we gave for $\alpha=0.5$  and a bigger
%value for $\alpha=\beta=0.6$, e.g. $10$.\\
%With this simple  model we get the right  value for the fresh  mixture
%and a likely  value for denser  mixture, in accord with the hypothesis
%of stationarity of the lower layer.
%\begin{obs}
%  This would not  be the only possible function  $\eta(\alpha)$ and it
%  is not one  of our aims to discuss how realistic  this behaviour is. 
%  Here we  needed just a  value of $\eta$,  not too far from  the real
%  one, for  any value of  the solid fraction  $\alpha$, to use  in the
%  numerical  simulations.  In  fact we  will  use only  the values  of
%  $\eta$  for   $\alpha\leq  0.5$,  for  the  upper   layer,  and  for
%  $\alpha=\beta$, for  the middle layer, considering  the viscosity of
%  the bottom layer high enough to keep it at rest.
%\end{obs}
%%% Local Variables: 
%%% mode: latex
%%% TeX-master: t
%%% End: 
%\input{final}                                                   
\section{The final equations} \label{sec:final}

Using the definition of $D_1$, $D_2$ in ~\eqref{eq:derivate}, dividing
~\eqref{eq:f=ma1}  by   $A_1$   and ~\eqref{eq:f=ma2}  by    $A_2$ and
subtracting the second equation from the first, we obtain:
\begin{equation}\label{eq:3-bifase}
  \left(A_2\rho_1+A_1\rho_2\right)\partial_tU+
  \left(A_2\rho_1U+A_1\rho_2V\right)\partial_xU=
  \rho_2\dot{Q}-\dfrac{A_2}{A_1}\tau_1S_1+
  \tau_2S_2
  \mp\tau_iS_1\left(\dfrac{A_2}{A_1}+1\right)
\end{equation}
which is  the third differential  equation of the  model (together with
~\eqref{eq:1-bifase}  ~\eqref{eq:2-bifase}),  in  the  three  unknowns
$\alpha$, $A_1$, $U$.  Equation ~\eqref{eq:V} gives
\[V=V(\alpha,A_1,U)\]
while $\beta$ is given.

We  can   easily rewrite   ~\eqref{eq:1-bifase}, ~\eqref{eq:2-bifase},
~\eqref{eq:3-bifase} in matrix  form, introducing $M$, $N$, $3\times3$
matrices and {\boldmath$\Omega$}, $\mbf F$, two column vectors:
\begin{equation}\label{eq:Omega-F}
\left.
  \begin{array}{ll}
    \text{{\boldmath$\Omega$}}=\left(\begin{array}{c}
        \alpha\\
        A_1\\
        U\\
      \end{array}\right) & \mbf F=\left(\begin{array}{c} 
        -\alpha\left(1-\dfrac{\alpha}{\beta}\right)\psi\\
        -A_1\dfrac{\alpha}{\beta}\psi\\
        f\\
      \end{array}\right)
  \end{array}
\right.\end{equation}
where 
\begin{equation}\label{eq:f}
  f=\rho_2\dot{Q}-\dfrac{A_2}{A_1}\tau_1S_1+
  \tau_2S_2
  \mp\tau_iS_1\left(\dfrac{A_2}{A_1}+1\right)
\end{equation}
\begin{equation}\label{eq:M-N}
\left.
  \begin{array}{ll}
    M=\left(
      \begin{array}{ccc}
        1 & 0 & 0\\
        0 & 1 & 0\\
        0 & 0 &  A_2\rho_1+A_1\rho_2\\
      \end{array}\right) &
    N=\left(
      \begin{array}{ccc}
        U & 0 & 0\\
        0 & U & A_1\\
        0 & 0 & A_2\rho_1U+A_1\rho_2V\\
      \end{array}\right)
  \end{array}\right.
\end{equation}
so that:
\begin{equation}\label{eq:sistema}
  M\partial_t\text{{\boldmath$\Omega$}}+N\partial_x\text{{\boldmath$\Omega$}}=\mbf F
\end{equation}
\begin{thm}[Well-posedness]\label{thm:hyp1}
  The problem ~\eqref{eq:sistema} with boundary-Cauchy data:
  \begin{equation} \label{eq:cauchy-data}
    \left\{
      \begin{array}{l}
        \text{{\boldmath$\Omega$}}(0,t)=\text{{\boldmath$\Omega$}}^0\; ;\\
        \text{{\boldmath$\Omega$}}(x,0)=\text{{\boldmath$\Omega$}}^0\; ;
      \end{array}
    \right.
  \end{equation}
  is well-posed.
\end{thm}
\begin{proof}
  We  begin by  showing  that ~\eqref{eq:sistema}  is hyperbolic  (see
\cite{couranthilbert},  \cite{garabedian},  \cite{jeffrey}).   $M$  is
invertible, so multiply ~\eqref{eq:sistema}  on both sides by $M^{-1}$
and   obtain  the  normal   form  of   hyperbolic  systems   of  PDE's
\begin{equation}\label{eq:normale}
\partial_t\text{{\boldmath$\Omega$}}+L\partial_x\text{{\boldmath$\Omega$}}=\mbf H \end{equation} where
now:  \begin{equation}\label{eq:L} L=M^{-1}N=\left( \begin{array}{ccc}
U    &    0     &0\\    0    &    U    &    A_1\\     0    &    0    &
\dfrac{A_2\rho_1U+A_1\rho_2V}{A_2\rho_1+A_1\rho_2}\\
\end{array}\right)   \end{equation}  and  \begin{equation}\label{eq:H}
\mbf          H=M^{-1}\mbf          F=\left(          \begin{array}{c}
-\alpha\left(1-\dfrac{\alpha}{\beta}\right)\psi\\
-A_1\dfrac{\alpha}{\beta}\psi\\ h\\ \end{array} \right) \end{equation}
where   \begin{equation}\label{eq:h}  h=\dfrac{1}{A_2\rho_1+A_1\rho_2}
\left(\rho_2\dot{Q}-\dfrac{A_2}{A_1}\tau_1S_1+\tau_2S_2
\mp\tau_iS_1\left(\dfrac{A_2}{A_1}+1\right)\right)  \end{equation}  As
it   can  be  easily   seen,  $L$   has  real   positive  eigenvalues:
\begin{equation}\label{eq:eigenvals}            \left\{\begin{array}{l}
\lambda_1,                                                \lambda_2=U\\
\lambda_3=\dfrac{A_2\rho_1U+A_1\rho_2V}{A_2\rho_1+A_1\rho_2}\\
\end{array}\right.  \end{equation} Moreover,  as the three eigenvalues
are  positive and  finite, the  boundary-Cauchy data  on the  two axes
($x=0$   ,   $t=0$),   are    given   on   time-like   segments   (see
\cite{couranthilbert}, ~\cite{jeffrey}), and the well-posedness of the
problem is guaranted.
\end{proof}
\begin{obs} \label{obs:ill-posed}
  Ill-posedness of  previous models formulated for similar non-steady,
  multiphase flow,(see \cite{liepmann})  had already  been pointed out
  in different  papers  (see \cite{cheng-wang-ling}, \cite{ramshaw-trapp},
  \cite{stewart}) where different methods to bypass ill-posedness were
  proposed, based on the introduction of surface tension terms. Such a
  procedure, which   would  be not  applicable to   our case,   is not
  necessary in our model which is consistently formulated as a correct
  evolution problem in a natural way.
\end{obs}
%%% Local Variables: 
%%% mode: latex
%%% TeX-master: t
%%% End: 
%\input{adim}
\subsection{Non-dimensional variables}

Before deriving the complete model with  the introduction of the third
layer, let us define a typical length-scale $L_c$  (e.g. the length of
the pipeline, between two pumping  stations) and a typical  time-scale
$t^0$  so to use non-dimensional  variables;  the  time-scale will  be
chosen  to be $\dfrac{L_c}{U^0}$, with $U^0$the  velocity of the fresh
mixture at the entrance of the pipe.  In particular  $L_c$ will be the
average   distance between two  pumping   stations ($\simeq 100 km$),
while $U^0$ will  be $2.0 ms^{-1}$ in  accord with the expected  total
flow rate ($\simeq  1424 m^3h^{-1}$, ($0.4 m^3s^{-1}$)), with  a section diameter of $0.5
m$.
\begin{equation}
  \begin{array}{ll}
    \tau=\dfrac{t}{t^0}{\text ,} & \xi=\dfrac{x}{L_c}
  \end{array}
\end{equation}
Now we can use non-dimensional unknowns:
\begin{equation}
  \begin{array}{ll}
    a_{1,2}=\dfrac{A_{1,2}}{A}, & v_{1,2}=\dfrac{(U,V)}{U^0}
  \end{array}
\end{equation}

We use also non-dimensional  densities, scaled with  the density of water
$\rho_w$:
\begin{equation}
  \delta_{1,2}=\dfrac{\rho_{1,2}}{\rho_w}\;,
\end{equation}
non-dimensional cross-sections perimeters, scaled by the perimeter of the
pipeline:
\begin{equation}
  \sigma_{1,2,i}=\dfrac{S_{1,2,i}}{S}\;,
\end{equation}
non-dimensional specific shear stresses
\begin{equation}
  \mu_{1,2,i}=\dfrac{\tau_{1,2,i}}{{\lambda_1}^0\rho_w{U^0}^2}
\end{equation}
where ${\lambda_1}^0$ is the  friction factor of the wall shear-stress
for the upper layer, evaluated at the boundary.

We  can  rewrite  the  system ~\eqref{eq:normale}  in a completely  non
dimensional form:
\begin{equation}\label{eq:adim2}
  \partial_\tau\mbf U+A\partial_\xi\mbf U=\mbf S
\end{equation}
where:
\begin{equation}\label{eq:U-S}
  \begin{array}{ll}
    \mbf U=\left(\begin{array}{c}
        \alpha\\
        a_1\\
        v_1\\
      \end{array}\right), &
    \mbf S=\left(\begin{array}{c}
        \alpha\left(1-\dfrac{\alpha}{\beta}\right)t^0\psi\\
        -\dfrac{\alpha}{\beta}a_1t^0\psi\\
        s\\
      \end{array}\right)
  \end{array}
\end{equation}
with
\begin{equation}
  s=\dfrac{1}{a_2\delta_1+a_1\delta_2}\left[{\lambda_1}^0\dfrac{2\pi RL_c}{A}
    \left(-\mu_1\dfrac{a_2}{a_1}\sigma_1+\mu_2\sigma_2\mp\mu_i\left(\dfrac{a_2}{a_1}+
        1\right)\sigma_i\right)\right]
\end{equation}
and
\begin{equation}\label{eq:A}
  A=\left(\begin{array}{ccc}
      v_1  & 0 & 0\\
      0 & v_1 & a_1\\
      0 & 0 & \dfrac{\delta_1a_2v_1+\delta_2a_1v_2}{\delta_1a_2+\delta_2a_1}\\
    \end{array}\right)
\end{equation}
\begin{obs}
  Here and  in the following  we considered constant the  total volume
  discharge along the pipe, i.e.
  \begin{equation}
    Q=AU^0
  \end{equation}
\end{obs}
%%% Local Variables: 
%%% mode: latex
%%% TeX-master: t
%%% End: 
%\input{threelayer}
\section{The three-layer flow model} \label{sec:threelayer}

Let  us  now   introduce  the  third  layer,  seen,   as  we  said  in
sec.~\ref{sec:introduction},  as a  stationary  deposit with  constant
solid  volume   fraction  $\gamma$,  and  let  us   write  the  volume
conservation   equations,   introducing   the   new   transfer   rates
$\overline\psi$,   $\overline\varphi$,  of   the   solid  and   liquid
components from layer 2 to layer 3.
\begin{figure}[htb]                                                    
  \begin{picture}(0,0)%
\includegraphics{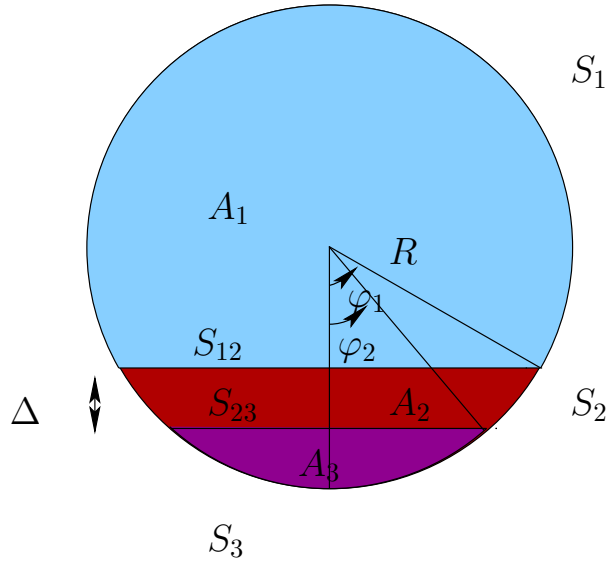}%
\end{picture}%
\setlength{\unitlength}{3947sp}%
\begingroup\makeatletter\ifx\SetFigFont\undefined%
\gdef\SetFigFont#1#2#3#4#5{%
  \reset@font\fontsize{#1}{#2pt}%
  \fontfamily{#3}\fontseries{#4}\fontshape{#5}%
  \selectfont}%
\fi\endgroup%
\begin{picture}(3534,3431)(1876,-2583)
\put(5396,364){\makebox(0,0)[lb]{\smash{\SetFigFont{14}{16.8}{\familydefault}{\mddefault}{\updefault}$S_1$}}}
\put(3021,-1347){\makebox(0,0)[lb]{\smash{\SetFigFont{14}{16.8}{\familydefault}{\mddefault}{\updefault}$S_{12}$}}}
\put(4256,-1727){\makebox(0,0)[lb]{\smash{\SetFigFont{14}{16.8}{\familydefault}{\mddefault}{\updefault}$A_2$}}}
\put(5396,-1727){\makebox(0,0)[lb]{\smash{\SetFigFont{14}{16.8}{\familydefault}{\mddefault}{\updefault}$S_2$}}}
\put(3115,-1727){\makebox(0,0)[lb]{\smash{\SetFigFont{14}{16.8}{\familydefault}{\mddefault}{\updefault}$S_{23}$}}}
\put(3686,-2108){\makebox(0,0)[lb]{\smash{\SetFigFont{14}{16.8}{\familydefault}{\mddefault}{\updefault}$A_3$}}}
\put(3115,-2583){\makebox(0,0)[lb]{\smash{\SetFigFont{14}{16.8}{\familydefault}{\mddefault}{\updefault}$S_3$}}}
\put(3115,-492){\makebox(0,0)[lb]{\smash{\SetFigFont{14}{16.8}{\familydefault}{\mddefault}{\updefault}$A_1$}}}
\put(4256,-777){\makebox(0,0)[lb]{\smash{\SetFigFont{14}{16.8}{\familydefault}{\mddefault}{\updefault}$R$}}}
\put(3935,-1349){\makebox(0,0)[lb]{\smash{\SetFigFont{14}{16.8}{\familydefault}{\mddefault}{\updefault}$\varphi_2$}}}
\put(1876,-1786){\makebox(0,0)[lb]{\smash{\SetFigFont{14}{16.8}{\familydefault}{\mddefault}{\updefault}$\Delta$}}}
\put(3995,-1045){\makebox(0,0)[lb]{\smash{\SetFigFont{14}{16.8}{\familydefault}{\mddefault}{\updefault}$\varphi_1$}}}
\end{picture}
\caption{Three-layer flow model, vertical section}                   
  \label{fig:3fasi} 
\end{figure}                    
Then, the volume conservation equations will be:
\begin{itemize}
\item LAYER 1-solid:
  \begin{equation}\label{eq:S1}
    \partial_t(\alpha A_1)+\partial_x(\alpha A_1U)=-\alpha A_1\psi    
  \end{equation}
\item LAYER 1-liquid:
  \begin{equation}\label{eq:L1}
    \partial_t\left[(1-\alpha)A_1\right]+\partial_x\left[(1-\alpha) 
      A_1U\right]=-(1-\alpha) A_1\varphi  
  \end{equation}
\item LAYER 2-solid
  \begin{equation} \label{eq:S2}
    \partial_t(\beta A_2)+\partial_x(\beta A_2V)=
    \alpha A_1\left(\psi-\o\psi\right)
  \end{equation}                                                     
\item LAYER 2-liquid
  \begin{equation} \label{eq:L2} 
    \partial_t\left[(1-\beta) A_2\right]+\partial_x\left[(1-\beta)
      A_2V\right]=(1-\alpha) A_1\left(\varphi-\o\varphi\right)
  \end{equation} 
\item LAYER 3-solid
  \begin{equation} \label{eq:S3}
    \partial_t(\gamma A_3)=\alpha A_1 \overline{\psi}
  \end{equation}
\item LAYER 3-liquid
  \begin{equation} \label{eq:L3}
    \partial_t[(1-\gamma)A_3]=(1-\alpha)A_1\overline{\varphi}
  \end{equation}                                                               
\end{itemize}                                                                 
\begin{obs}
  The reason why the quantity  of matter passing from the middle layer
  to the stationary  deposit is proportional to $\alpha  A_1$, and not
  to  the corresponding  factor of  the middle  layer $\beta  A_2$, in
  ~\eqref{eq:S3},  is due  to the  fact that  we suppose  the thickness
  $\Delta$ of  the middle layer constant  . From this  it follows that
  the transfer  rate across  the intermediate layer  is driven  by the
  trasfer of matter between the upper and the middle layer.
  
  In this way,  the mass transfer to the  stationary deposit stops, as
  well as the one from the  upper to middle layer, only when the upper
  layer  is  completely  empty  of  solid ($\alpha=0$).  This  is  the
  condition of  complete phase separation and  consequent steady flow,
  which is reached asimptotically by  our system, although in a region
  of  not practical  interest,  since  it is  far  from the  operating
  conditions of a plant.
\end{obs}
Now we have
\begin{equation}\label{eq:Atot}
  A=A_1+A_2+A_3
\end{equation}
and
\begin{equation}\label{eq:beta-gamma-const}
  \beta,\gamma=const.
\end{equation}
Summing up equations from ~\eqref{eq:S1} to ~\eqref{eq:L3}, we still have
the total volume conservation:
\begin{equation}\label{eq:Vol-tot}
  \partial_x\left(A_1U+A_2V\right)=0
\end{equation}
which means again:
\begin{equation} \label{eq:V-3}
  V=\dfrac{Q-A_1U}{A_2}\;,\; \text{if}\;A_2>0
\end{equation} 
Dividing   ~\eqref{eq:S3}  by     $\gamma$   and  ~\eqref{eq:L3}    by
$(1-\gamma)$, after  using ~\eqref{eq:beta-gamma-const}, and equating
the results, we get:
\begin{equation}\label{eq:phisign}
  \overline\varphi=\dfrac{\alpha}{1-\alpha}\dfrac{1-\gamma}{\gamma}
  \overline\psi
\end{equation}
which replaces ~\eqref{eq:varphi(psi)}.

Dividing ~\eqref{eq:S2} by  $\beta$ and ~\eqref{eq:L2} by $(1-\beta)$,
and equating the results, we have now:
\begin{equation} \label{eq:phi-psi}
  \varphi=\dfrac{1-\beta}{\beta}\dfrac{\alpha}{1-\alpha}\psi+
  \dfrac{\alpha}{1-\alpha}\left(\dfrac{1-\gamma}{\gamma}-
    \dfrac{1-\beta}{\beta}\right)\overline{\psi}
\end{equation}
Note that, being the thickness of the middle layer constant, we can express
\begin{equation}\label{eq:A2(A3)}
  A_2=A_2(A_3,\Delta)
\end{equation}
from which it follows:
\begin{equation}\label{eq:derivA2}
  \dfrac{\partial A_2}{\partial(x,t)}=
  \dfrac{\partial A_2}{\partial A_3}\dfrac{\partial A_3}{\partial(x,t)}
\end{equation}
\begin{obs}[On the thickness of the middle and bottom layer]
  To simplify  the equation and  without losing physical  meaning, we
  will suppose that  the thickness of the sedimentation  layers can be
  considered small enough to have (see Fig. ~\ref{fig:3fasi}):
  \begin{equation}\label{eq:angles}
    sin\varphi_i\simeq\varphi_i-\dfrac{{\varphi_i}^3}{6}\;\text{,}\; i=1,2
  \end{equation}
  With this hypothesis, we will have:
  \begin{equation}\label{eq:A_2}
    A_2\simeq\dfrac{2}{3}R^2(\varphi_1^3-\varphi_2^3)
  \end{equation}
  and
  \begin{equation}\label{eq:A_3}
    A_3\simeq\dfrac{2}{3}R^2\varphi_2^3
  \end{equation}
  while $\varphi_1$ and $\varphi_2$ will be related by:
  \begin{equation}\label{eq:delta_const}
    \Delta\simeq\dfrac{1}{2}R(\varphi_1^2-\varphi_2^2)
  \end{equation}
  This assumption  can be  made without any  loss of  physical meaning
  since  we know  that the  presence of  the stationary  deposit  is a
  damage for  the correct use  of the plant.  Therefore,  in practical
  applications  in plant,  its thickness  must be  kept below  a small
  fraction of the pipe radius.
\end{obs}
From
  ~\eqref{eq:Atot} we get
\begin{equation} \label{eq:areaA_3}
  \dfrac{\partial A_3}{\partial(x,t)}=-\dfrac{\partial A_1}{\partial(x,t)}
  -\dfrac{\partial A_2}{\partial(x,t)}
\end{equation}
Using ~\eqref{eq:derivA2} in ~\eqref{eq:areaA_3}, we have then
\begin{equation} \label{eq:derivA3}
  \dfrac{\partial A_3}{\partial(x,t)}=
  -\dfrac{1}{1+\dfrac{\partial A_2}{\partial A_3}}
  \dfrac{\partial A_1}{\partial(x,t)}
\end{equation}
Now substitute ~\eqref{eq:derivA3} in ~\eqref{eq:S3}:
\begin{equation} \label{eq:psisign}
  \overline{\psi}=-\dfrac{\gamma}{\alpha}\dfrac{1}{1+\dfrac{\partial A_2}
    {\partial A_3}}\dfrac{\partial_t A_1}{A_1}
\end{equation}
which completes     the  relation between     $\varphi$  and $\psi$ in
~\eqref{eq:phi-psi}.
\begin{obs}
  We will actually suppose  that the transfer  of matter from the
  middle layer to stationary deposit  starts only when the thickness of
  the middle layer has reached  the constant  value $\Delta$. We  will
  suppose in particular that:
  \[\overline{\psi}=-\dfrac{\gamma}{\alpha}\dfrac{1}{1+\dfrac{\partial A_2}
    {\partial     A_3}}\dfrac{\partial_t   A_1}{A_1}\Gamma(h)\]  where
  $\Gamma(h)$  is the step  function of the  thickness $h$  of the middle
  layer:
  \begin{equation}
    \Gamma(h)=\left\{\begin{array}{ll}
        0 & \text{if}\,\, h\leq\Delta\\
        1 & \text{if}\,\, h > \Delta\\
      \end{array}\right.
  \end{equation}
\end{obs}
Using ~\eqref{eq:beta-gamma-const} we can rewrite  ~\eqref{eq:S2} as:
\begin{equation} \label{eq:S2nuova}
  \partial_tA_2+\partial_x(A_2V)=
  \dfrac{\alpha}{\beta} A_1\left(\psi-\o\psi\right)
\end{equation}                                                     
and   using  ~\eqref{eq:derivA2}  together  with  ~\eqref{eq:derivA3},
~\eqref{eq:psisign} and   the definition of  $V$, ~\eqref{eq:V-3}, we
get:
\begin{equation} \label{eq:3fasi-2}
  \dfrac{\dfrac{\partial A_2}{\partial A_3}+\dfrac{\gamma}{\beta}}
  {1+\dfrac{\partial A_2}{\partial A_3}}                             
  \partial_t A_1+\partial_x(A_1U)=
  -A_1\dfrac{\alpha}{\beta}\psi
\end{equation}
Now we divide ~\eqref{eq:S1} by $\alpha$ and subtract ~\eqref{eq:3fasi-2}
from the resulting equation:
\begin{equation}
  \dfrac{A_1}{\alpha}\partial_t\alpha+\left(1-
    \dfrac{\dfrac{\partial A_2}{\partial A_3}+\dfrac{\gamma}{\beta}}
    {1+\dfrac{\partial A_2}{\partial A_3}}\right)\partial_tA_1+
  \dfrac{A_1U}{\alpha}\partial_x\alpha=-A_1
  \left(1-\dfrac{\alpha}{\beta}\right)\psi
\end{equation}
that we can rewrite:
\begin{equation}\label{eq:3fasi-1}
  \partial_t\alpha-\alpha\dfrac
  {\dfrac{\gamma}{\beta}-1}{1+\dfrac{\partial A_2}{\partial A_3}}
  \dfrac{\partial_tA_1}{A_1}+U\partial_x\alpha=
  -\alpha\left(1-\dfrac{\alpha}{\beta}\right)\psi
\end{equation}
We will take ~\eqref{eq:3fasi-1}  and  ~\eqref{eq:3fasi-2} as the  the
first two equations of the final model in the unknowns $\alpha,A_1,U$.
\begin{obs}
  Equation  ~\eqref{eq:3fasi-1}  can   be rewritten, using    the usual
  definition of the derivatives along the flow,
  \begin{equation}\label{eq:derivalpha}
    D_1\alpha=-\alpha\left(1-\dfrac{\alpha}{\beta}\right)\psi+
    \alpha\dfrac{\dfrac{\gamma}{\beta}-1}{1+\dfrac{\partial A_2}
      {\partial A_3}}\dfrac{\partial_tA_1}{A_1}
  \end{equation}
  and, since  $\partial_tA_1\leq  0$, $\alpha<\beta$  and $\beta<\gamma$,
  $\alpha$ is strictly decreasing along the flow.
\end{obs}
%%% Local Variables: 
%%% mode: latex
%%% TeX-master: t
%%% End: 
%\input{momentum}
\subsection{Momentum balance}\label{sec:momentum}

In   the  following   we   will   use  the   same   notations  as   in
sec.~\ref{sec:2fasi} with  the changes due to the  introduction of the
stationary deposit, as described in Fig.~\ref{fig:3fasi}.

The momentum  exchange will  be now between  the upper and  the middle
layer, with  exactly the same terms  as before and a  term of momentum
loss of  the middle  layer, in favour  of the stationary  deposit. The
momentum transferred  to the stationary deposit  is completely absorbed
by  the pipe  wall.  With this  in mind,  let  us write  down the  two
equations of momentum balance for the layers in motion.
\begin{itemize} 
\item LAYER 1:
  \begin{equation} \label{eq:Mom1}
    D_1(A_1\rho_1U)=A_1G-\tau_1S_1\mp\tau_{12}S_{12}+UD_1(\rho_1A_1)
  \end{equation}
\item LAYER 2:
  \begin{equation} \label{eq:Mom2}
    D_2(A_2\rho_2V)=A_2G-\tau_2S_2\pm\tau_{12}S_{12}-\tau_{23}S_{23}
    -\rho_2UD_2A_1-\rho_2VD_2A_3
  \end{equation}
\end{itemize}
%\begin{obs}
%  Note the differences in the stress terms index, due to the change of
%  the section perimeters    after the introduction of the   stationary
%  deposit (see fig.~\ref{fig:3fasi}).
%\end{obs}
Using ~\eqref{eq:derivA3}, ~\eqref{eq:V-3} in ~\eqref{eq:Mom2}, dividing
~\eqref{eq:Mom1}   by  $A_1$  and    ~\eqref{eq:Mom2}  by $A_2$,   and
subtracting the first from the second, we get the third equation of the
complete model:
\begin{eqnarray}\label{eq:3fasi-3}
  \lefteqn{\dfrac{\rho_2V}{1+\dfrac{\partial A_2}{\partial A_3}}
    \partial_t A_1+\left(A_2\rho_1+A_1\rho_2\right)\partial_tU+
    \dfrac{\rho_2V^2}{1+\dfrac{\partial A_2}{\partial A_3}}
    \partial_xA_1+\left(A_2\rho_1U+A_1\rho_2V\right)
    \partial_xU=}\\\nonumber
  & &=-\dfrac{A_2}{A_1}\tau_1S_1+\tau_2S_2\mp\tau_{12}S_{12}
  \left(\dfrac{A_2}{A_1}+1\right)+\tau_{23}S_{23}
\end{eqnarray}
Before going further with  the calculations,  let us reduce  equations
~\eqref{eq:3fasi-1},     ~\eqref{eq:3fasi-2},  ~\eqref{eq:3fasi-3}  to
non-dimensional form, with the same scale factors as before:
\begin{equation} \label{eq:3Fasi-1}
  \partial_\tau\alpha-\dfrac{\dfrac{\gamma}{\beta}-1}
  {\dfrac{\partial a_2}{\partial a_3}+1}\dfrac{\alpha}{a_1}
  \partial_\tau a_1+v_1\partial_\xi\alpha=-\alpha
  \left(1-\dfrac{\alpha}{\beta}\right)t^0\psi
\end{equation}
\begin{equation} \label{eq:3Fasi-2}
  \dfrac{\dfrac{\partial a_2}{\partial a_3}+\dfrac{\gamma}{\beta}}
  {1+\dfrac{\partial a_2}{\partial a_3}}\partial_\tau a_1+
  \partial_\xi\left(a_1v_1\right)=-a_1\dfrac{\alpha}{\beta}t^0\psi
\end{equation}
\begin{eqnarray} \label{eq:3Fasi-3}
  \lefteqn{\dfrac{\dfrac{\gamma}{\beta}-1}{1+\dfrac{\partial a_2}{\partial a_3}}
    \dfrac{\delta_2v_2}{\dfrac{\gamma}{\beta}-1}\partial_\tau a_1+
    \left(a_2\delta_1+a_1\delta_2\right)\partial_\tau v_1+}\\ \nonumber
  & &  +\dfrac{\dfrac{\gamma}{\beta}-1}{1+\dfrac{\partial a_2}{\partial a_3}}
  \dfrac{\delta_2{v_2}^2}{\dfrac{\gamma}{\beta}-1}\partial_\xi a_1+
  \left(a_2\delta_1v_1+a_1\delta_2v_2\right)\partial_\xi v_1=\\\nonumber
  & & ={\lambda_1}^0\dfrac{2L_c}{R}\left(-\dfrac{a_2}{a_1}\mu_1\sigma_1
    +\mu_2\sigma_2\mp\left(\dfrac{a_2}{a_1}+1\right)\mu_{12}\sigma_{12}+
    \mu_{23}\sigma_{23}\right)
\end{eqnarray}
The boundary-Cauchy data, will be a constant vector:
\begin{equation}\label{eq:U0}
  \mbf U^0=\left(\begin{array}{c}
      \alpha^0\\
      1\\
      1\\
    \end{array}\right)
\end{equation}
\begin{thm}[Well-posedness]\label{thm:hyp2}
  The    system   of  PDE    ~\eqref{eq:3Fasi-1}, ~\eqref{eq:3Fasi-2},
  ~\eqref{eq:3Fasi-3} with boundary-Cauchy data:
  \begin{equation}\label{eq:b-C-data}
    \left\{\begin{array}{l}
        \mbf U(0,\tau)=\mbf U^0\;;\\
        \mbf U(\xi,0)=\mbf U^0\;;\\
      \end{array}\right.
  \end{equation}
  is  well-posed   in  the  range   of  validity  of   the  assumption
  ~\eqref{eq:angles}.
\end{thm}
\begin{proof}
  The system of equations        ~\eqref{eq:3Fasi-1}  ~\eqref{eq:3Fasi-2}
  ~\eqref{eq:3Fasi-3} can be reduced again to normal matrix form:
  \begin{equation}\label{eq:Norm}
    \partial_\tau\mbf U+ L\partial_\xi\mbf U=\mbf S
  \end{equation}
  where, as usual:
  \begin{equation}
    \left.
      \begin{array}{ll}
        \mbf U=\left(
          \begin{array}{c}
            \alpha\\
            a_1\\
            v_1\\
          \end{array}\right) &
        \mbf S=\left(\begin{array}{c}
            -\alpha\left(1-\dfrac{\dfrac{\partial a_2}{\partial a_3}+1}
              {\dfrac{\partial a_2}{\partial a_3}+\dfrac{\gamma}{\beta}}
              \dfrac{\alpha}{\beta}\right)t^0\psi\\
            -\dfrac{\dfrac{\partial a_2}{\partial a_3}+1}
            {\dfrac{\partial a_2}{\partial a_3}+\dfrac{\gamma}{\beta}}
            a_1\dfrac{\alpha}{\beta}t^0\psi\\
            s\\
          \end{array}\right)
      \end{array}\right.
  \end{equation}
  with
  \begin{eqnarray}\label{eq:s}
    \lefteqn{s=\dfrac{1}{\dfrac{\partial a_2}{\partial a_3}
        +\dfrac{\gamma}{\beta}}\dfrac{\alpha}{\beta}
      \dfrac{a_1\delta_2 v_2}{a_2\delta_1+a_1\delta_2}t^0\psi+}\\\nonumber
    & & +\dfrac{{\lambda_1}^0}{a_2\delta_1+a_1\delta_2}
    \dfrac{2L_c}{R}\left(-\dfrac{a_2}{a_1}\mu_1\sigma_1
      +\mu_2\sigma_2\mp\left(\dfrac{a_2}{a_1}+1\right)\mu_{12}\sigma_{12}+
      \mu_{23}\sigma_{23}\right)
  \end{eqnarray}
  and where the matrix $L$ is now:
  \begin{equation}\label{eq:LL}
    L=\left(\begin{array}{ccc}
        v_1 & \dfrac{\dfrac{\gamma}{\beta}-1}{\dfrac
          {\partial a_2}{\partial a_3}+\dfrac{\gamma}{\beta}}\dfrac{\alpha v_1}{a_1}
        & \dfrac{\dfrac{\gamma}{\beta}-1}{\dfrac
          {\partial a_2}{\partial a_3}+\dfrac{\gamma}{\beta}}\alpha\\
        0 & \dfrac{1+\dfrac{\partial a_2}{\partial a_3}}{\dfrac{\gamma}{\beta}+
          \dfrac{\partial a_2}{\partial a_3}}v_1
        & \dfrac{1+\dfrac{\partial a_2}{\partial a_3}}{\dfrac{\gamma}{\beta}+
          \dfrac{\partial a_2}{\partial a_3}}a_1\\
        0 & l_{32} & l_{33}\\
      \end{array}\right)
  \end{equation}
  with:
  \begin{equation}\label{eq:k}
    k=\dfrac{\delta_2v_2}{a_2\delta_1+a_1\delta_2}\dfrac{1}
    {\dfrac{\gamma}{\beta}-1}\; ,
  \end{equation}
  \[\left\{
    \begin{array}{l}
      l_{32}=-\dfrac{\dfrac{\gamma}{\beta}-1}
      {\dfrac{\partial a_2}{\partial a_3}+1}k
      \left(v_1-v_2\dfrac{\dfrac{\partial a_2}{\partial a_3}+
          \dfrac{\gamma}{\beta}}{\dfrac{\partial a_2}{\partial a_3}+1}\right)\\
      l_{33}=\dfrac{a_2\delta_1v_1+a_1\delta_2v_2}{a_2\delta_1+a_1\delta_2}
      -\dfrac{\dfrac{\gamma}{\beta}-1}{\dfrac{\partial a_2}
        {\partial a_3}+1}a_1k
    \end{array}
  \right.\]
  It is easy to see that, defined a function
  \begin{equation}\label{eq:eps}
    \epsilon=\dfrac{\dfrac{\gamma}{\beta}-1}
    {\dfrac{\partial a_2}{\partial a_3}+\dfrac{\gamma}{\beta}}\; ,
  \end{equation}
  the matrix $L$ can be rewritten
  \begin{equation} \label{eq:L=A+C}
    L=A+\epsilon C
  \end{equation}
  where $A$ is the matrix of the two-layer model
  \begin{equation}\label{eq:AA}
    A=\left(\begin{array}{ccc}
        v_1  & 0 & 0\\
        0 & v_1 & a_1\\
        0 & 0 & \dfrac{\delta_1a_2v_1+\delta_2a_1v_2}{\delta_1a_2+
          \delta_2a_1}\\
      \end{array}\right)
  \end{equation}
  and $C$ is 
  \begin{equation}\label{eq:CC}
    C=\left(\begin{array}{ccc}
        0 & \dfrac{\alpha v_1}{a_1} & \alpha\\
        0 & -v_1 & - a_1\\
        0 & -k\left(v_1-v_2\dfrac{\dfrac{\partial a_2}{\partial a_3}+
            \dfrac{\gamma}{\beta}}
          {\dfrac{\partial a_2}{\partial a_3}+1}\right) & 
        -a_1k\\
      \end{array}
    \right)
  \end{equation}
  so  (see  Remark   \ref{obs:epsilon}),  for  small  $\epsilon$,  the
  introduction  of the  stationary  deposit  can be  seen  as a  small
  perturbation of the two-layer flow model.  
  
  We want  to show that  the equation ~\eqref{eq:Norm} is  hyperbolic. 
  The first  eigenvalue of $L$ is  easily seen to be  $v_1$, the other
  two are the zeros of the polynomial:
\begin{equation}\label{eq:p_L}
  p_L(\lambda)=(l_{22}-\lambda)(l_{33}-\lambda)-l_{32}l_{23}
\end{equation} To show that the  eigenvalues are real, we show that:
\begin{equation}\label{eq:Delta_L}
  {\Delta_L}^2=(l_{22}+l_{33})^2-4(l_{22}l_{33}-l_{23}l_{32})\geq 0
\end{equation} 
in the hypothesis of small stationary deposit. 

This can be rewritten:
\begin{equation}
  {\Delta_L}^2=(l_{22}-l_{33})^2+4l_{23}l_{32}\geq 0
\end{equation}
which means:
\begin{equation}\label{eq:prima}
  {\Delta_L}^2=\left[(1-\epsilon)v_1-{\lambda_3}^A+\epsilon a_1k\right]^2
  -4\epsilon a_1k\left(v_1(1-\epsilon)-v_2\right)
\end{equation}
where we used:
\begin{equation}
  {\lambda_3}^A=\dfrac{a_1\delta_2v_2+a_2\delta_1v_1}
  {a_1\delta_2+a_2\delta_1}
\end{equation}
the third eigenvalue of the matrix $A$.

Let:
\begin{equation}\label{eq:omega_def}
\omega=\dfrac{a_1\delta_2}{a_1\delta_2+a_2\delta_1}
\end{equation}
and
\begin{equation}\label{eq:Gamma_def}
\Gamma=\dfrac{\beta}{\gamma-\beta}
\end{equation}
Now we can rewrite eq. ~\eqref{eq:prima}:
\begin{equation}\label{eq:parabola}
{\Delta_L}^2=\omega^2\left[\left(1-\Gamma\epsilon\right)^2+
4\dfrac{\epsilon\Gamma}{\omega}\right]v_2^2
-2\omega v_1\left[\left(1-\Gamma\epsilon\right)(\omega-\epsilon)+
2\epsilon\Gamma(1-\epsilon)\right]v_2+v_1^2(\omega-\epsilon)^2
\end{equation}
Calling 
\begin{equation}\label{eq:epsilon_}
\overline\epsilon=\Gamma\epsilon \text{,}
\end{equation}
~\eqref{eq:parabola} can be rewritten in the more compact way:
\begin{equation}\label{eq:parabola2}
{\Delta_L}^2=v_1^2(\omega-\epsilon)^2\left\{\left(\dfrac{\omega}{\omega-\epsilon}\right)^2\left[(1-\o\epsilon)^2+4\dfrac{\o\epsilon}{\omega}\right]u^2-2\left(\dfrac{\omega}{\omega-\epsilon}\right)\left[(1-\o\epsilon)+\dfrac{2\o\epsilon(1-\epsilon)}{\omega-\epsilon}\right]u+1\right\}
\end{equation}
where
\begin{equation}\label{eq:v2/v1}
u=\dfrac{v_2}{v_1}
\end{equation}
Therefore, the zeros of $\Delta_L^2$ correspond  to the zeros of a
polynomial in $u$ of the form:
\[\left(\dfrac{\omega}{\omega-\epsilon}\right)^2a
u^2-2\left(\dfrac{\omega}{\omega-\epsilon}\right)bu+1\]
that has solutions for
\[u_{1,2}=\left(\dfrac{\omega-\epsilon}{\omega}\right)\dfrac{b\pm\sqrt{b^2-a}}{a}\]
After some algebra we will see that these solutions are in fact out of
the region in wich $v_2$ varies. 

Neglecting the $\epsilon^3$ terms (see Rem. ~\ref{obs:epsilon}), we find that:
\begin{equation}\label{eq:b^2-ac}
b^2-a \simeq 4\dfrac{\o\epsilon^2}{\omega^2}\left[\left(1-\omega\right)\left(\dfrac{1}{\Gamma}+1\right)\right]
\end{equation} 
So, neglecting the $\epsilon^2$ terms in $b$ and $a$, we end up with:
\begin{equation}\label{eq:sol_u}
  u_{1,2}\simeq \left(\dfrac{\omega-\epsilon}{\omega}\right)
  \dfrac{1-\o\epsilon+\dfrac{2\o\epsilon}{\omega}\pm
    \dfrac{2\o\epsilon}{\omega}\sqrt{\left(1-\omega\right)
      \left(\dfrac{1}{\Gamma}+1\right)}}{1+2\o\epsilon\left(\dfrac{2}{\omega}-1\right)}
\end{equation}
where  $\epsilon,\o\epsilon<<1$   (see  Rem.   \ref{obs:epsilon}).

From the definition of $\omega$, we see that:
\begin{equation}\label{eq:1-omega}
  1-\omega=\dfrac{a_2\delta_1}{a_2\delta_1+a_1\delta_2}
\end{equation}
which is of order $\varphi_1^3-\varphi_2^3$ (see eq. ~\eqref{eq:A_2}),
so that  in the  expression ~\eqref{eq:sol_u}, the  difference between
the  two zeros  $u_1$ and  $u_2$ is  small in  comparison  with $u_1$,
$u_2$, which are both close to
\begin{equation}\label{eq:u_approx}
  \o{u}\sim 1-\o\epsilon\left(1+2\dfrac{a_2\delta_1}{a_1\delta_2}\right)\;\text ,
\end{equation}
the difference being of the order of $\varphi_2\varphi_1^{2/3}$.

From the last  we can actually appreciate that $\o u$  has to be close
to $1$, to  make $\Delta_L^2$ negative, conditions that  will never be
reached   in    the   hypothesis   ~\eqref{eq:angles}    (see   Remark
\ref{obs:thm}).
\end{proof}

\begin{obs}[On the validity of Theorem \ref{thm:hyp2}]\label{obs:thm}
  It is clear that the  interval in which $\Delta_L^2(u)$ is negative,
  is very  close to $u=1$, conditions  in which we  are not interested
  (see Rem.  \ref{obs:a3} ), corresponding to large phase separation
  (see  Rem.  \ref{obs:epsilon}), negligible in practical applications.\\
  From asymptotic study (see App.  \ref{sec:App}) we know in fact that
  $v_2$, after a jump-like behaviour for $\xi <<1$, starts increasing,
  pushed  by  the  faster,  less  dense, upper  phase,  but  remaining
  considerably smaller than $v_1$, also increasing.
\end{obs}

\begin{obs}[On $\epsilon$] \label{obs:epsilon}
  It is easy to see that, for $a_3<<1$,
  \begin{equation}\label{eq:epsilon}
    \epsilon\simeq\dfrac{\varphi_2}{\varphi_2+\Gamma
      \left(\varphi_2^2+2\dfrac{\Delta}{R}\right)^{1/2}}
  \end{equation} with  $\varphi_2$, the  angle subtended by  $a_3$ (see
  Fig.   \ref{fig:3fasi}).  \\Therefore  $\epsilon  <<1$  if
  \begin{equation}\label{eq:lim_phi2}
    \dfrac{\varphi_2}{\Gamma\left(2\dfrac{\Delta}{R}\right)^{1/2}}<<1\;\text ,
  \end{equation}
  which defines what we mean by a thin stationary deposit.
\end{obs}                                                                   
\begin{obs}\label{obs:a3}
  The case  in which  $a_3$ is not  small, like,  e.g., in the  case of
  large phase  separation, should be treated differently  and will not
  be considered  in this model.   For example in problems  of restart,
  the  pressure gradient,  that we  considered  the same  for the  two
  layers  in  motion,  is  different  at different  levels,  during  a
  transient phase.
\end{obs}
%%% Local Variables: 
%%% mode: latex
%%% TeX-master: t
%%% End: 
%\input{Cauchy-bound}
\section{On the boundary-Cauchy data}

In sections ~\ref{sec:final} and ~\ref{sec:momentum} we introduced the
boundary-Cauchy data.  In  particular, in theorems ~\ref{thm:hyp1} and
~\ref{thm:hyp2}  we  imposed  constant  data on  both  axes  ($\xi=0$,
$\tau=0$).

It has  to be noted  that this choice  was not the  only possible.  We
chose that at the entrance the  condition of the  mixture is, at every
time  $t$, the same as  it was in the  whole pipe at time $t=0$. 

%The latter condition (eq.~\eqref{eq:condiz}) is rather artificial, and
%is, in fact, a way of avoiding restart problems, that, as we said, are
%considerably different.

%Note also that this choice does not influence in  any way the proof of
%well posedness.   
%It does influence   the  regularity of  solution see
%theorem 2.1, pag. 71, in ~\cite{jeffrey}. 

Note that  the regularity of  the boundary-Cauchy data  influences the
regularity of solution (see  ~\cite{jeffrey}, Theorem 2.1, pag 71). In
our case, being the functions  $L$ and $S$ in ~\eqref{eq:Norm} regular
and choosing  constant the boundary-Cauchy data, the  solution will be
regular too.

Our choice will be therefore:
\begin{equation} \label{eq:b-C}
  \mbf U^0=\left(
    \begin{array}{c}
      \alpha^0\\
      1\\
      1\\
    \end{array}
  \right)
\end{equation}
This corresponds to the choice we  made of having at time $t=0$ and at
any $t$, at  the entering cross section, the pipe  full of matter with
solid fraction $\alpha^0$ and velocity $U^0$, namely:
\[\left.
  \begin{array}{ccc}
    \alpha^0=0.5 & A_1^0=A & U^0=2 m/s\\
  \end{array}
\right.\]

\subsection{On $V^0$}

Some remarks have to be made for the  value of the velocity of
the middle layer at the entrance. 

From  eq. ~\eqref{eq:V} we  see that in  fact $V$ is defined only for
$A_2\neq 0$.  However this is not true in our conditions, for $x=0$. We
can calculate the value $V^0$ as:
\begin{equation} \label{eq:lim}
  V^0=\lim_{x\rightarrow 0}\dfrac{Q(t)-A_1U}{A_2}
\end{equation}
To do this we have  to take into  account the  different terms in the
source   function in  eq.  ~\eqref{eq:f=ma1}  and ~\eqref{eq:f=ma2},
evaluated in the vicinity  of $x=0$.   All  the terms of  the right-hand
side vanish, linearly with $A_2$ or with $S_i$, as $A_2\rightarrow 0$,
apart from the  wall shear-stress term  $\tau_2S_2$  vanishing only in
turbulent regime.  In this case, $V^0$ turns out to be equal to $U^0$,
so the  second layer  appears  with the   same  velocity as the  fresh
mixture.

On the other hand, choosing  the laminar regime, we get $V^0=0$, which
is however consistent with the model only if $\psi$ is taken no longer
constant, but a function of $x$, vanishig at $x=0$.

We put in  the Appendix the study of the  asymptotics near the origin,
in which we distinguish the different cases of flow regime.

In the  numerical simulations we chose  the turbulent-turbulent regime
because it fits more naturally the model with $\psi$ constant.

A possible  genaralization could be  the introduction of  a transition
from one regime to another in the first segment of the pipe.
%%% Local Variables: 
%%% mode: latex
%%% TeX-master: t
%%% End: 
%\input{simulations}
\section{The numerical results} To evaluate the solution of
eq.~\eqref{eq:Norm}  we  elaborated  a  numerical code,  based  on  an
explicit approximation method . Eq. ~\eqref{eq:Norm} is discretised as
\begin{equation}\label{eq:linear}
  u[i+1][x_j]-u[i][x_j]=\Delta t
  \left\{L[i][x_j]\dfrac{(u[i][x_j]-u[i][x_{j-1}])}
  {\Delta x}\right\}+H[i][x_j]
\end{equation}
where  we indicate with $u[i][x_j]$   the solution of ~\eqref{eq:Norm}
evaluted at   $i-th$ time step,  in  the  $j-th$ cell   of the spatial
domain.

%This method, with a judicious choice  of the dimension of the cells of
%the  space-time grid,  gives  good confidence  of  stability and  quite
%acceptable approximation,  and allows  the simulations to  be run  on a
%normal PC.
This  finite  differences method  of  approximation  explicit in  time
(Euler explicit) with a judicious choice of the dimension of the cells
of the space-time  grid, gives good confidence of  stability and quite
acceptable approximation,  and allows the  simulations to be run  on a
normal  PC.  In  the  numerical code  we  inserted a  control for  the
dimension   of  the  space-time   grid,  in   order  to   satisfy  the
Courant-Friedrichs-Lewy\footnote{The condition states that,
given the  eigenvalues of the characteristic  matrix, $\lambda_p$, and
being $h$ (along $x$) and $k$  (along $t$) the dimensions of the cells
in the space-time grid, necessary  condition for the convergence of an
approximate method for PDE systems, is that:
\[\left|\dfrac{\lambda_p k}{h}\right|\leq 1\]
In the case of the complete system (see sec.~\ref{sec:threelayer}), we
calculate numerically the eigenvalues of the characteristic matrix and
we reject the time steps at  which the CFL condition is not satisfied. 
For  a detailed discussion  on the  convergence of  this approximation
method, see ~\cite{isaacson}.}  condition (see ~\cite{leveque}) during
the whole run.

The solution of equation  ~\eqref{eq:Norm} can be plotted at different
time steps as a function  of the normalized distance from the entrance
$\xi$ ($=x/L_c$).  We plot,  in the graphs below different quantities:
$\alpha$, $v_1$ and $1-a_1$, solutions of ~\eqref{eq:Norm}, as well as
some  important  ones, related  to  the  solution vector  ($\mbf{U}$),
namely $v_2$, $a_3$.

\begin{figure}[htb]                                                           
\begin{picture}(0,0)%
\includegraphics{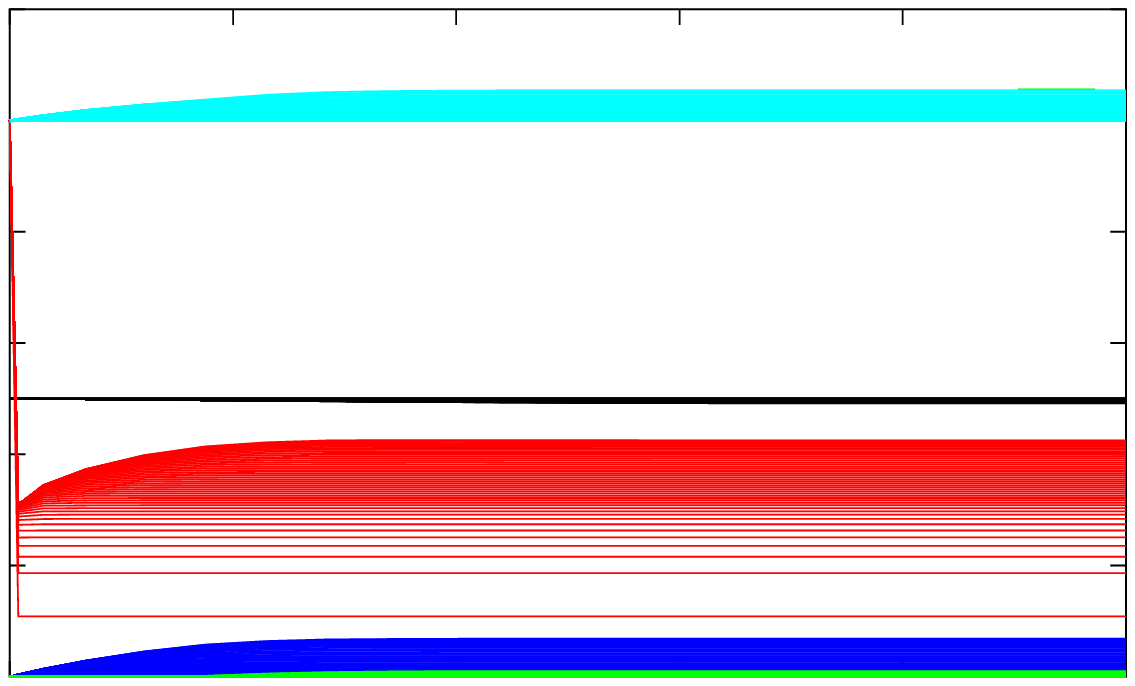}%
\end{picture}%
\setlength{\unitlength}{3947sp}%
\begingroup\makeatletter\ifx\SetFigFont\undefined%
\gdef\SetFigFont#1#2#3#4#5{%
  \reset@font\fontsize{#1}{#2pt}%
  \fontfamily{#3}\fontseries{#4}\fontshape{#5}%
  \selectfont}%
\fi\endgroup%
\begin{picture}(5567,3720)(659,-3286)
\put(3226,-3286){\makebox(0,0)[lb]{\smash{\SetFigFont{12}{14.4}{\familydefault}{\mddefault}{\updefault}
$\xi$
}}}
\put(6226,-2761){\makebox(0,0)[lb]{\smash{\SetFigFont{10}{12.0}{\familydefault}{\mddefault}{\updefault}
$a_3$
}}}
\put(659,-2875){\makebox(0,0)[rb]{\smash{\SetFigFont{10}{12.0}{\familydefault}{\mddefault}{\updefault}
0
}}}
\put(659,-2341){\makebox(0,0)[rb]{\smash{\SetFigFont{10}{12.0}{\familydefault}{\mddefault}{\updefault}
0.2
}}}
\put(659,-1807){\makebox(0,0)[rb]{\smash{\SetFigFont{10}{12.0}{\familydefault}{\mddefault}{\updefault}
0.4
}}}
\put(659,-1273){\makebox(0,0)[rb]{\smash{\SetFigFont{10}{12.0}{\familydefault}{\mddefault}{\updefault}
0.6
}}}
\put(659,-739){\makebox(0,0)[rb]{\smash{\SetFigFont{10}{12.0}{\familydefault}{\mddefault}{\updefault}
0.8
}}}
\put(659,-205){\makebox(0,0)[rb]{\smash{\SetFigFont{10}{12.0}{\familydefault}{\mddefault}{\updefault}
1
}}}
\put(659,329){\makebox(0,0)[rb]{\smash{\SetFigFont{10}{12.0}{\familydefault}{\mddefault}{\updefault}
1.2
}}}
\put(733,-2999){\makebox(0,0)[b]{\smash{\SetFigFont{10}{12.0}{\familydefault}{\mddefault}{\updefault}
0
}}}
\put(1805,-2999){\makebox(0,0)[b]{\smash{\SetFigFont{10}{12.0}{\familydefault}{\mddefault}{\updefault}
0.2
}}}
\put(2876,-2999){\makebox(0,0)[b]{\smash{\SetFigFont{10}{12.0}{\familydefault}{\mddefault}{\updefault}
0.4
}}}
\put(3948,-2999){\makebox(0,0)[b]{\smash{\SetFigFont{10}{12.0}{\familydefault}{\mddefault}{\updefault}
0.6
}}}
\put(5019,-2999){\makebox(0,0)[b]{\smash{\SetFigFont{10}{12.0}{\familydefault}{\mddefault}{\updefault}
0.8
}}}
\put(6091,-2999){\makebox(0,0)[b]{\smash{\SetFigFont{10}{12.0}{\familydefault}{\mddefault}{\updefault}
1
}}}
\put(4351,-1336){\makebox(0,0)[rb]{\smash{\SetFigFont{10}{12.0}{\familydefault}{\mddefault}{\updefault}
$\alpha$
}}}
\put(4276, 89){\makebox(0,0)[rb]{\smash{\SetFigFont{10}{12.0}{\familydefault}{\mddefault}{\updefault}
$v_1$
}}}
\put(3001,-2461){\makebox(0,0)[rb]{\smash{\SetFigFont{10}{12.0}{\familydefault}{\mddefault}{\updefault}
$1-a_1$
}}}
\put(2326,-1636){\makebox(0,0)[rb]{\smash{\SetFigFont{10}{12.0}{\familydefault}{\mddefault}{\updefault}
$v_2$
}}}
\end{picture}
\caption{Numerical simulation  of  the three
    layers  flow model  for  half a  characteristic time  ($t^0=200000
    sec.$), $\Delta=5 cm$} \label{fig:orange_005_col}
\end{figure}                                                                  
In Fig.  ~\ref{fig:orange_005_col} we  show the result of a simulation
of  the  evolution of  the  system  in 100000  sec.,  as  we said,  in
turbulent-turbulent  regime. We  see the  formation of  the stationary
deposit after the  achievement of the maximum thickness  of the middle
layer.  Here we  chose $U^0=0.5 m/s$ and $L_c=100  km$, with a maximum
thickness $\Delta=5 cm$.

\begin{figure}[htb]                                                           
\begin{picture}(0,0)%
\includegraphics{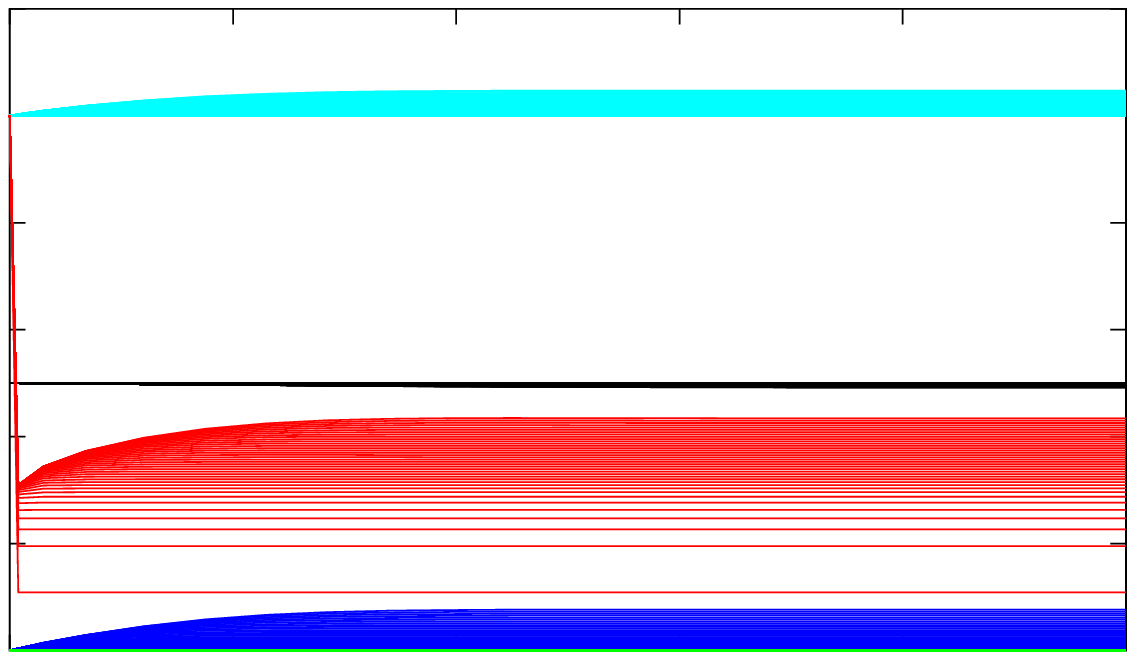}%
\end{picture}%
\setlength{\unitlength}{3947sp}%
\begingroup\makeatletter\ifx\SetFigFont\undefined%
\gdef\SetFigFont#1#2#3#4#5{%
  \reset@font\fontsize{#1}{#2pt}%
  \fontfamily{#3}\fontseries{#4}\fontshape{#5}%
  \selectfont}%
\fi\endgroup%
\begin{picture}(5444,3495)(659,-3061)
\put(659,-2751){\makebox(0,0)[rb]{\smash{\SetFigFont{10}{12.0}{\familydefault}{\mddefault}{\updefault}
0
}}}
\put(659,-2238){\makebox(0,0)[rb]{\smash{\SetFigFont{10}{12.0}{\familydefault}{\mddefault}{\updefault}
0.2
}}}
\put(659,-1724){\makebox(0,0)[rb]{\smash{\SetFigFont{10}{12.0}{\familydefault}{\mddefault}{\updefault}
0.4
}}}
\put(659,-1211){\makebox(0,0)[rb]{\smash{\SetFigFont{10}{12.0}{\familydefault}{\mddefault}{\updefault}
0.6
}}}
\put(659,-698){\makebox(0,0)[rb]{\smash{\SetFigFont{10}{12.0}{\familydefault}{\mddefault}{\updefault}
0.8
}}}
\put(659,-184){\makebox(0,0)[rb]{\smash{\SetFigFont{10}{12.0}{\familydefault}{\mddefault}{\updefault}
1
}}}
\put(659,329){\makebox(0,0)[rb]{\smash{\SetFigFont{10}{12.0}{\familydefault}{\mddefault}{\updefault}
1.2
}}}
\put(733,-2875){\makebox(0,0)[b]{\smash{\SetFigFont{10}{12.0}{\familydefault}{\mddefault}{\updefault}
0
}}}
\put(1805,-2875){\makebox(0,0)[b]{\smash{\SetFigFont{10}{12.0}{\familydefault}{\mddefault}{\updefault}
0.2
}}}
\put(2876,-2875){\makebox(0,0)[b]{\smash{\SetFigFont{10}{12.0}{\familydefault}{\mddefault}{\updefault}
0.4
}}}
\put(3948,-2875){\makebox(0,0)[b]{\smash{\SetFigFont{10}{12.0}{\familydefault}{\mddefault}{\updefault}
0.6
}}}
\put(5019,-2875){\makebox(0,0)[b]{\smash{\SetFigFont{10}{12.0}{\familydefault}{\mddefault}{\updefault}
0.8
}}}
\put(6091,-2875){\makebox(0,0)[b]{\smash{\SetFigFont{10}{12.0}{\familydefault}{\mddefault}{\updefault}
1
}}}
\put(3412,-3061){\makebox(0,0)[b]{\smash{\SetFigFont{12}{14.4}{\familydefault}{\mddefault}{\updefault}
$\xi$
}}}
\put(4801,-1336){\makebox(0,0)[rb]{\smash{\SetFigFont{10}{12.0}{\familydefault}{\mddefault}{\updefault}
$\alpha$
}}}
\put(3676, 14){\makebox(0,0)[rb]{\smash{\SetFigFont{10}{12.0}{\familydefault}{\mddefault}{\updefault}
$v_1$
}}}
\put(5551,-2386){\makebox(0,0)[rb]{\smash{\SetFigFont{10}{12.0}{\familydefault}{\mddefault}{\updefault}
$1-a_1$
}}}
\put(1351,-1636){\makebox(0,0)[rb]{\smash{\SetFigFont{10}{12.0}{\familydefault}{\mddefault}{\updefault}
$v_2$
}}}
\end{picture}
\caption{Numerical simulation  of  the three
   layers  flow  model for  half  a  characteristic time  ($t^0=200000
   sec.$), $\Delta=8 cm$} \label{fig:orange_008_col}
\end{figure}

In similar conditions, but with  a maximum thickness $\Delta=8 cm$, we
do not see any stationary deposit (see Fig.~\ref{fig:orange_008_col}).

\begin{figure}[htb]                                                           
\begin{picture}(0,0)%
\includegraphics{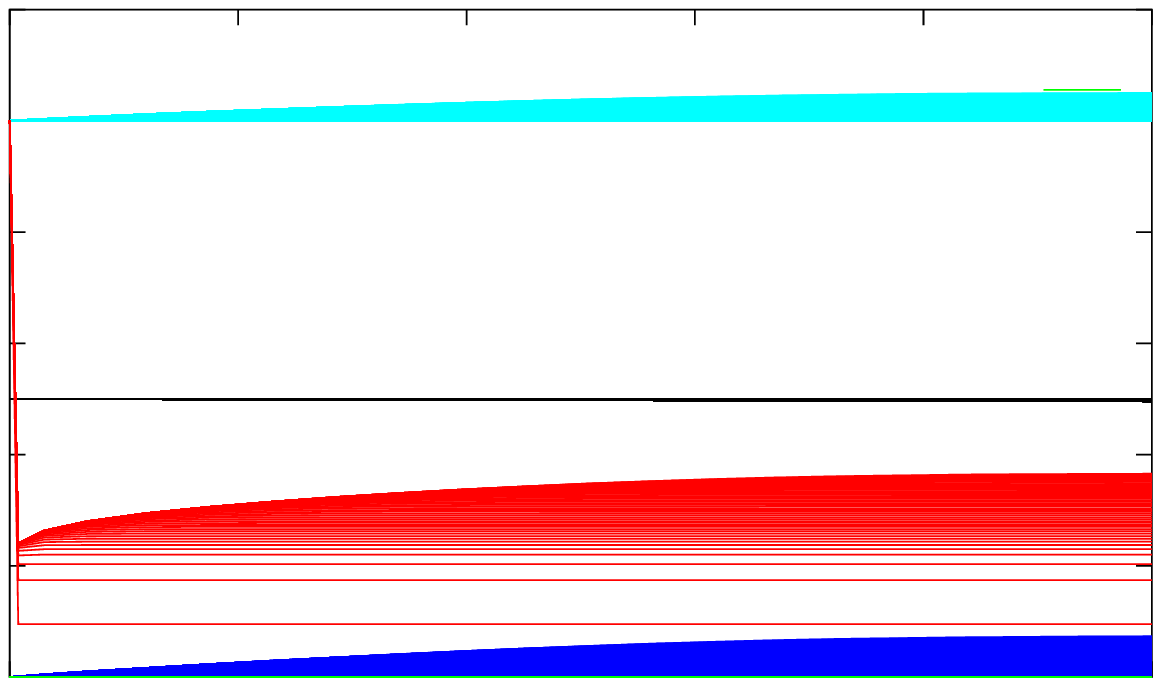}%
\end{picture}%
\setlength{\unitlength}{3947sp}%
\begingroup\makeatletter\ifx\SetFigFont\undefined%
\gdef\SetFigFont#1#2#3#4#5{%
  \reset@font\fontsize{#1}{#2pt}%
  \fontfamily{#3}\fontseries{#4}\fontshape{#5}%
  \selectfont}%
\fi\endgroup%
\begin{picture}(5568,3645)(597,-3211)
\put(3376,-3211){\makebox(0,0)[lb]{\smash{\SetFigFont{12}{14.4}{\familydefault}{\mddefault}{\updefault}
$\xi$
}}}
\put(597,-2875){\makebox(0,0)[rb]{\smash{\SetFigFont{10}{12.0}{\familydefault}{\mddefault}{\updefault}
0
}}}
\put(597,-2341){\makebox(0,0)[rb]{\smash{\SetFigFont{10}{12.0}{\familydefault}{\mddefault}{\updefault}
0.2
}}}
\put(597,-1807){\makebox(0,0)[rb]{\smash{\SetFigFont{10}{12.0}{\familydefault}{\mddefault}{\updefault}
0.4
}}}
\put(597,-1273){\makebox(0,0)[rb]{\smash{\SetFigFont{10}{12.0}{\familydefault}{\mddefault}{\updefault}
0.6
}}}
\put(597,-739){\makebox(0,0)[rb]{\smash{\SetFigFont{10}{12.0}{\familydefault}{\mddefault}{\updefault}
0.8
}}}
\put(597,-205){\makebox(0,0)[rb]{\smash{\SetFigFont{10}{12.0}{\familydefault}{\mddefault}{\updefault}
1
}}}
\put(597,329){\makebox(0,0)[rb]{\smash{\SetFigFont{10}{12.0}{\familydefault}{\mddefault}{\updefault}
1.2
}}}
\put(671,-2999){\makebox(0,0)[b]{\smash{\SetFigFont{10}{12.0}{\familydefault}{\mddefault}{\updefault}
0
}}}
\put(1767,-2999){\makebox(0,0)[b]{\smash{\SetFigFont{10}{12.0}{\familydefault}{\mddefault}{\updefault}
0.2
}}}
\put(2864,-2999){\makebox(0,0)[b]{\smash{\SetFigFont{10}{12.0}{\familydefault}{\mddefault}{\updefault}
0.4
}}}
\put(3960,-2999){\makebox(0,0)[b]{\smash{\SetFigFont{10}{12.0}{\familydefault}{\mddefault}{\updefault}
0.6
}}}
\put(5057,-2999){\makebox(0,0)[b]{\smash{\SetFigFont{10}{12.0}{\familydefault}{\mddefault}{\updefault}
0.8
}}}
\put(6153,-2999){\makebox(0,0)[b]{\smash{\SetFigFont{10}{12.0}{\familydefault}{\mddefault}{\updefault}
1
}}}
\put(3301, 89){\makebox(0,0)[rb]{\smash{\SetFigFont{10}{12.0}{\familydefault}{\mddefault}{\updefault}
$v_1$
}}}
\put(1801,-1786){\makebox(0,0)[rb]{\smash{\SetFigFont{10}{12.0}{\familydefault}{\mddefault}{\updefault}
$v_2$
}}}
\put(4276,-1336){\makebox(0,0)[rb]{\smash{\SetFigFont{10}{12.0}{\familydefault}{\mddefault}{\updefault}
$\alpha$
}}}
\put(5026,-2536){\makebox(0,0)[rb]{\smash{\SetFigFont{10}{12.0}{\familydefault}{\mddefault}{\updefault}
$1-a_1$
}}}
\end{picture}
\caption{Numerical simulation  of the  three layers
  flow  model   for  two  characteristic   times  ($t^0=50000  sec.$),
  $\Delta=8 cm$} \label{fig:100_2_2}
\end{figure}

\begin{figure}[htb]                                                           
\begin{picture}(0,0)%
\includegraphics{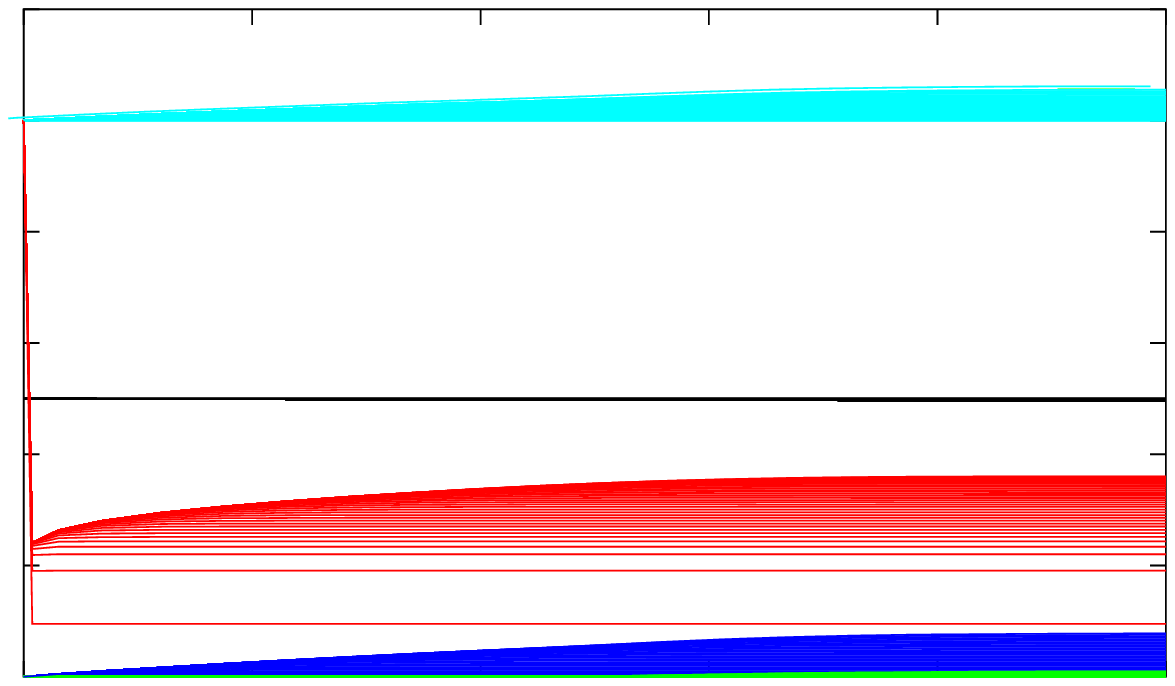}%
\end{picture}%
\setlength{\unitlength}{3947sp}%
\begingroup\makeatletter\ifx\SetFigFont\undefined%
\gdef\SetFigFont#1#2#3#4#5{%
  \reset@font\fontsize{#1}{#2pt}%
  \fontfamily{#3}\fontseries{#4}\fontshape{#5}%
  \selectfont}%
\fi\endgroup%
\begin{picture}(6016,3720)(585,-3286)
\put(3151,-3286){\makebox(0,0)[lb]{\smash{\SetFigFont{12}{14.4}{\familydefault}{\mddefault}{\updefault}
$\xi$
}}}
\put(597,-2875){\makebox(0,0)[rb]{\smash{\SetFigFont{10}{12.0}{\familydefault}{\mddefault}{\updefault}
0
}}}
\put(597,-2341){\makebox(0,0)[rb]{\smash{\SetFigFont{10}{12.0}{\familydefault}{\mddefault}{\updefault}
0.2
}}}
\put(597,-1807){\makebox(0,0)[rb]{\smash{\SetFigFont{10}{12.0}{\familydefault}{\mddefault}{\updefault}
0.4
}}}
\put(597,-1273){\makebox(0,0)[rb]{\smash{\SetFigFont{10}{12.0}{\familydefault}{\mddefault}{\updefault}
0.6
}}}
\put(597,-739){\makebox(0,0)[rb]{\smash{\SetFigFont{10}{12.0}{\familydefault}{\mddefault}{\updefault}
0.8
}}}
\put(597,-205){\makebox(0,0)[rb]{\smash{\SetFigFont{10}{12.0}{\familydefault}{\mddefault}{\updefault}
1
}}}
\put(597,329){\makebox(0,0)[rb]{\smash{\SetFigFont{10}{12.0}{\familydefault}{\mddefault}{\updefault}
1.2
}}}
\put(671,-2999){\makebox(0,0)[b]{\smash{\SetFigFont{10}{12.0}{\familydefault}{\mddefault}{\updefault}
0
}}}
\put(1767,-2999){\makebox(0,0)[b]{\smash{\SetFigFont{10}{12.0}{\familydefault}{\mddefault}{\updefault}
0.2
}}}
\put(2864,-2999){\makebox(0,0)[b]{\smash{\SetFigFont{10}{12.0}{\familydefault}{\mddefault}{\updefault}
0.4
}}}
\put(3960,-2999){\makebox(0,0)[b]{\smash{\SetFigFont{10}{12.0}{\familydefault}{\mddefault}{\updefault}
0.6
}}}
\put(5057,-2999){\makebox(0,0)[b]{\smash{\SetFigFont{10}{12.0}{\familydefault}{\mddefault}{\updefault}
0.8
}}}
\put(6153,-2999){\makebox(0,0)[b]{\smash{\SetFigFont{10}{12.0}{\familydefault}{\mddefault}{\updefault}
1
}}}
\put(4426, 89){\makebox(0,0)[rb]{\smash{\SetFigFont{10}{12.0}{\familydefault}{\mddefault}{\updefault}
$v_1$
}}}
\put(2401,-1786){\makebox(0,0)[rb]{\smash{\SetFigFont{10}{12.0}{\familydefault}{\mddefault}{\updefault}
$v_2$
}}}
\put(5476,-1336){\makebox(0,0)[rb]{\smash{\SetFigFont{10}{12.0}{\familydefault}{\mddefault}{\updefault}
$\alpha$
}}}
\put(6601,-2761){\makebox(0,0)[rb]{\smash{\SetFigFont{10}{12.0}{\familydefault}{\mddefault}{\updefault}
$a_3$
}}}
\put(4501,-2461){\makebox(0,0)[rb]{\smash{\SetFigFont{10}{12.0}{\familydefault}{\mddefault}{\updefault}
$1-a_1$
}}}
\end{picture}
\caption{Numerical simulation  of  the three
  layers flow  model for two characteristic  times ($t^0=50000 sec.$),
  $\Delta=5 cm$} 
\label{fig:100_2_0_05}
\end{figure}

With the initial velocity  $U^0=2   m/s$ and $\Delta=8cm$, again,   no
stationary  deposit is   present  on  a   lenght   of $100   km$  (see
Fig.~\ref{fig:100_2_2}), while the   stationary deposit is  visible in
similar   conditions,       but  with         $\Delta=5cm$        (See
Fig. ~\ref{fig:100_2_0_05}).
From the above  simulations we see the dependence  of the deposit upon
the  thickness, $\Delta$,  of  the  middle layer  and  upon the  input
velocity $U^0$.  The  stationary deposit is, as we  expect, thinner in
the  case of  a larger  input  velocity and  disappears completely  in
figures  ~\ref{fig:orange_008_col}  and  ~\ref{fig:100_2_2} for  large
value of $\Delta$.

Note the jump-like  behaviour of $v_2$ near the  origin which simulates
the real behaviour $v_2\sim x^{0.2}$ (see App.~\ref{sec:App}).

We  have also  the  convergence,  observed for  bigger  values of  the
lenght-scale (out of the physical lenght of $100 km$) to an asymptotic
solution of  steady flow, corresponding  to the condition  of complete
phase separation. Our model converges  then to the steady models found
in     the    literature,    in     ~\cite{bifase2},    ~\cite{shook},
~\cite{shook-book}.
%%% Local Variables: 
%%% mode: latex
%%% TeX-master: t
%%% End: 
%\input{aknowledgements}
\section{Acknowledgements}
This project  was suggested by  Snamprogetti (Fano, Italy),  a leading
company  in the  field  of fuels  pipelining.   I am  thankful to  the
researching  staff   of  Snamprogetti,  for  providing   most  of  the
bibliographic material,  and for  the substantial contribution  to the
progress of this work.  I thank also Dr.  A.Mancini, from Universit\`a
di  Milano, for  his constant  assistance  in the  development of  the
numerical simulations and Prof.  A.Fasano from Universit\`a di Firenze
for his advise.
%%% Local Variables: 
%%% mode: latex
%%% TeX-master: t
%%% End: 
\vfill\eject                                                                 
\begin{appendix} \section{Asymptotic behaviour} \label{sec:App} 

We try here  to evaluate the behaviour of the  solutions of the system
of PDE' s we obtained, in the vicinity of the origin.

Here we  neglect the  dependence of the  solution upon  $t$, situation
that, as we  observed in the development of  the numerical simulations,
can  be  actually  reached  in  a  time  of  the  same  order  of  the
characteristic time $t^0$.

Let  us  suppose  then  that,  for small  $x$,  the  solution  vector
$\mbf\Omega$ has the expansion:
\begin{equation}\label{eq:sviluppo}
  \begin{array}{cc}
    A_1=A-kx^m-k_0x & A_2=kx^m+k_0x\\
    U=U^0+h_1x^{m_1}+{h_1}^0x &  V=V^0+h_2x^{m_2}+{h_2}^0x \\
  \end{array}
\end{equation}
with $m,m_1,m_2<1$.
  
From eq. ~\eqref{eq:2-bifase}, we get
\begin{equation}\label{eq:1-staz}
  \partial_x(A_1U)\simeq-\dfrac{\alpha^0}{\beta}A\psi
\end{equation}
From the conservation of total volume:
\begin{equation}\label{eq:cons}
  \partial_x(A_1U)=-\partial_x(A_2V)
\end{equation}
Using ~\eqref{eq:sviluppo}  in ~\eqref{eq:1-staz}, equating  the terms
of the same power, we get:
\begin{equation}\label{eq:condiz-1}
  \begin{array}{c}
    m\neq m_1\left\{
      \begin{array}{l}
        A{h_1}^0-k_0U^0=-A\dfrac{\alpha^0}{\beta}\psi\\
        h_1=k=0\\
      \end{array}\right.\\
    m=m_1\left\{
      \begin{array}{l}
        A{h_1}^0-k_0U^0=-A\dfrac{\alpha^0}{\beta}\psi\\
        Ah_1-U^0k=0\\
      \end{array}\right.\\
  \end{array}
\end{equation}
For $m+m_1>1$, while the case $m+m_1<1$ is not possible.

If, in particular, 
\begin{equation}
  m+m_1=1
\end{equation}
these conditions  change a little  in the case $m=m_1$,  while nothing
changes in the other cases:
\begin{equation}\label{eq:condiz-11}
  \begin{array}{c}
    m=m_1\left\{
      \begin{array}{l}
        -kh_1+A{h_1}^0-k_0U^0=-A\dfrac
        {\alpha^0}{\beta}\psi\\
        Ah_1-U^0k=0\\
      \end{array}\right.\\
  \end{array}
\end{equation}
Using now ~\eqref{eq:cons} in the same way:
\begin{equation}\label{eq:condiz-2}
  \begin{array}{l}
    m+m_2\neq 1\left\{
      \begin{array}{l}
        k_0V^0=A\dfrac{\alpha^0}{\beta}\psi\\
        kV^0=0\\
      \end{array}\right.\\
    m+m_2=1\left\{
      \begin{array}{l}
        kh_2+k_0V^0=A\dfrac{\alpha^0}{\beta}\psi\\
        kV^0=0\\
      \end{array}\right.\\
  \end{array}
\end{equation}
Note that,  in the  case $m+m_2\neq 1$,  it cannot be  $V^0=0$, while,
with $m+m_2=1$ both, $V^0=0$ and $V^0\neq 0$, are acceptable.

We  can  do  the same  with  the  two  equations of  momentum  balance
~\eqref{eq:f=ma1} ~\eqref{eq:f=ma2}; from the first we get:
\begin{equation}\label{eq:mom_approx-1}
  {\rho_1}^0AU^0\left(m_1h_1x^{m_1-1}+{h_1}^0\right)\simeq
  AG-S{\tau_1}^0-\tau_iS_i
\end{equation}
Then it has to be:
\begin{equation}
  h_1=0
\end{equation}
from    which    it     follows    (see    ~\eqref{eq:condiz-1}    and
~\eqref{eq:condiz-11})
\begin{equation}
  k=0
\end{equation}
in all the cases.

Moreover, as $\tau_iS_i$ is small near the origin,
\begin{equation}\label{eq:h_10}
  {h_1}^0=\dfrac{AG-S\tau_1^0}{{\rho_1}^0AU^0}
\end{equation}
Then the behaviour of  $U$ around the  origin depends  on the
pressure gradient.

From ~\eqref{eq:f=ma2} we get:
\begin{eqnarray}\label{eq:mom_approx-2}\nonumber
  \lefteqn{\rho_2k_0x(V^0+h_2^0x)(h_2^0+m_2h_2x^{m_2-1})\simeq}\\\nonumber
  & &\simeq k_0xG-\tau_2S_2+\tau_iS_i+\rho_2
  \left(U^0-V^0-h_2x^{m_2}+(h_1^0-h_2^0)x\right)(V^0+h_2x^{m_2}+h_2^0x)k_0\\
\end{eqnarray}
All the terms vanish in $x=0$, apart from
\[\rho_2(U^0-V^0)V^0k_0\]
then, it has to be either
\[V^0=0\]
or
\[V^0=U^0\]

In the case $V^0=U^0$, equating the linear terms we get also:
\begin{equation}\label{eq:h_20}
  h_2^0=\dfrac{G}{2\rho_2U^0}+\dfrac{h_1^0}{2}
\end{equation}
and,    as    $k$    is zero,   from    ~\eqref{eq:condiz-1}    and
~\eqref{eq:condiz-11} we have 
\begin{equation}\label{eq:k0}
  k_0=\dfrac{A}{U^0}\left(h_1^0+\dfrac{\alpha^0}{\beta}\psi\right)
\end{equation}
Now  let  us  call  $\varphi$  the  angle  subtended  by  $A_2$,  (see
fig.~\ref{fig:2fasi}).

It can be seen that, for small $\varphi$, 
\begin{equation}\label{eq:tau2}
  \tau_2S_2\simeq\omega_2V^{2-n}\varphi^{1-2n}
\end{equation}
where  $\omega_2$  is  a  positive  constant, and  $n$  is  the  power
appearing  in the  friction factor,  (see  sec.~\ref{sec:stress}),
depending on the flow regime.

Note that, to prevent this  term from becoming singular in the origin,
we need ${V\rightarrow 0}$ as ${x\rightarrow 0}$, if $n>\dfrac{1}{2}$.

We can rewrite ~\eqref{eq:tau2}
\begin{equation}\label{eq:tau22}
  \tau_2S_2\simeq{\omega_2}^\prime V^{2-n}x^{\frac{1-2n}{3}}
\end{equation}
where 
\[{\omega_2}^\prime=\left(\dfrac{3}{4}\dfrac{k_0}{R^2}\right)^{\frac{1}{3}}\omega_2\]
In the case $n\leq\dfrac{1}{2}$, it has to be then,
\[m_2=\dfrac{1-2n}{3}\]
from which it follows that:
\begin{equation}\label{eq:h_2}
  h_2=-\dfrac{{\omega_2}^\prime {U^0}^{1-n}}{\rho_2k_0(m_2+1)}
\end{equation}
So   we   have  $A_{1,2}$   and   $U$,   linear   in  $x$,   $V\simeq
U^0+h_2^0x+h_2x^{\frac{1-2n}{3}}$,  with  $h_2$  negative.  

The  case  $V^0=0$ is  actually  not  consistent  with the  assumption
$\psi=const.$

Let us see  how we have to  modify it in order to  make this behaviour
(suggested by the choice of laminar regime, $n=1$) acceptable.
%If we  now impose $V^0=0$,  as it would  happen in the  laminar regime
%($n=1$), we have to change a  little our hypothesis, to avoid that our
%equations become singular in the origin.

The dominant terms are,
\begin{itemize}
\item on the left-hand side:
  \[\rho_2k_0{h_2}^0h_2x^{m_2}\]
\item on the right-hand side:
  \[\left.\begin{array}{c}
      \tau_2S_2\simeq\omega_2^{\prime}h_2x^{m_2-\frac{1}{3}}\\
      \tau_iS_i\simeq\omega_i{U^0}^2x^{1+\frac{1}{3}}\\
      \rho_2U^0k_0h_2x^{m_2}\\
      \rho_2k_0{h_2}^2x^{2m_2}\\
    \end{array}\right.\]
\end{itemize}
Then, if $h_2\neq 0$, the $\tau_2S_2$ term  is balanced only if
$m_2=\dfrac{2}{3}$. This requires also:
\begin{equation}\label{eq:h2-lam}
  h_2=\dfrac{\omega_i}{{\omega_2}^\prime{U^0}^2}
\end{equation}
\begin{equation}\label{eq:h20-lam}
  h_2^0=U^0
\end{equation}
We said also:
\[\partial_x(A_2V)\simeq A\dfrac{\alpha^0}{\beta}\psi\]
which means
\begin{equation}\label{eq:singolare}
  k_0h_2(m_2+1)x^{m_2}+2k_0{h_2}^0x\simeq A\dfrac{\alpha^0}{\beta}\psi
\end{equation}
The matching above requires a  different choice for $\psi$. For example
we could say that,
\begin{equation}\label{eq:psi-nova}
  \psi(x)=\left\{\begin{array}{ll}
      \psi^0\dfrac{x}{x_0} & x<x_0\\
      \\
      \psi^0            & x\geq x_0\\
    \end{array}\right.
\end{equation}
In this way it has to be $h_2=0$ and
\begin{equation}\label{eq:h_20-nova}
  h_2^0=\dfrac{A\psi^0}{2k_0x_0}\dfrac{\alpha^0}{\beta}
\end{equation}
and also $V$ is linear around the origin; otherwise, if $h_2\neq 0$, we
must have:
\begin{equation}
  \psi\simeq\psi^0x^{m_2}
\end{equation}
and:
\begin{equation}
  h_2=\dfrac{A\dfrac{\alpha^0}{\beta}\psi^0}{k_0(m_2+1)}
\end{equation}
which is consistent with ~\eqref{eq:h2-lam}, only if we take:
\begin{equation}
  \psi^0=\dfrac{\omega_i}{{\omega_2}^\prime{U^0}^2}
  \dfrac{k_0(m_2+1)}{A\dfrac{\alpha^0}{\beta}}
\end{equation}
%With this choice of the  function $\psi(x)$ the behaviour of the other
%functions  would  change, as  now  $m_2=\dfrac{2}{3}$, while  choosing
%$\psi$  linear in $x$,  the solution  would not  change much the other
%variables. 
Now in ~\eqref{eq:condiz-1} we would have
\begin{equation}\label{h10-nova}
  k_0=\dfrac{Ah_1^0}{U^0}
\end{equation} 
and $\alpha,A_1,U$, would stay linear.
\end{appendix}
%%% Local Variables: 
%%% mode: latex
%%% TeX-master: t
%%% End: 
%\input{conclusion_App} 
\vfill\eject                                                                 
%\nocite{*}                                                                   
\bibliographystyle{plain} 
\bibliography{referenze}                                                     
\end{document}